\documentclass[letterpaper,11pt]{report}

\pdfoutput=1

\usepackage{amsmath, amsthm, amssymb}
\usepackage[dissertation]{USCthesis}
\usepackage{cite} 
\usepackage{url}  
\usepackage{ifthen}  
\usepackage{multicol}   
\usepackage[utf8]{inputenc} 
\usepackage{amsthm}
\usepackage{graphics}
\usepackage{graphicx}
\usepackage{epsfig}

\usepackage{longtable}
\usepackage[font=normalsize,labelfont=bf,labelsep=space]{caption}

\usepackage{comment}

\usepackage{algorithm}
\usepackage{algorithmic}

\theoremstyle{plain}

\theoremstyle{definition}

\begin{document}

\title{\doublespacing{INTEGRATIVE ANALYSIS OF \\ GENE EXPRESSION AND PHENOTYPE DATA}}
\author{Min Xu}
\submitdate{August 2009}

\universityname{UNIVERSITY OF SOUTHERN CALIFORNIA}
\schoolname{GRADUATE SCHOOL} \degree{DOCTOR OF PHILOSOPHY}
\majorfield{COMPUTATIONAL BIOLOGY}
\copyrightyear{2009}
\committee{Xianghong J. Zhou & (Chair person)\\*
               Fengzhu Sun \\*
               Gareth James & (Outside Member)}

\begin{preface}
\prefacesection{Dedication}
\begin{center}
{To Luge and Shiyou}
\end{center}

\prefacesection{Acknowledgements}

I would first like to express my appreciation to my adviser, Professor Xianghong Jasmine Zhou, for her insight, guidance, and support. Over the last few years,
her encouragement and commitment have been inspiring, and I am honored to take my place as an alumnus of her lab. I also want to thank Professor Gareth James,
Professor Fengzhu Sun, Professor Michael Waterman and Professor Donald Arnold for serving in either my Ph.D. candidacy or dissertation committees. I want to
thank Dr. Wenyuan Li, Professor Gareth James, Professor Louxin Zhang, Dr. Mike Mehan, Qiang Song, Dr. Juan Nunez-Iglesias and Dr. Ming-Chih Kao for their
cooperation, as well as Dr. Chao Cheng for manuscript reviews. My thanks also go to all the graduate students and postdoctoral fellows from our Computational
Biology Program for their friendship during my stay here.  I wish them great success in their studies and future careers.

I wish to extend my gratitude to other colleagues and friends, including Professor Frank Alber, Dr. Jianjun Hu, Dr. Jim Liu, Dr. Xiting Yan, Dr. Huanying Ge,
Dr. Kangyu Zhang, Dr. Zhidong Tu, Dr. Xiaotu Ma, Dr. Quansong Ruan, Dr. Li Wang, Dr. Weihong Xu, and Dr. Shihua Zhang, for their help during the course of my
studies.

Finally, I especially thank my wife, Luge Yang, for her dedication to our family and my father, Shiyou Wu, for his  encouragement and support. This
dissertation is dedicated to them.

\tableofcontents
\listoftables
\listoffigures

\prefacesection{Abstract}
The linking genotype to phenotype is the fundamental aim of modern genetics. We focus on study of links between gene expression data and phenotype data through
integrative analysis. In this work, we propose three approaches.

1) The inherent complexity of phenotypes makes high-throughput phenotype profiling a very difficult and laborious process. We propose a method of automated
multi-dimensional profiling which uses gene expression similarity. Large-scale analysis of more than 500 gene expression datasets show that our method can
provide robust profiling that reveals different phenotypic aspects of samples. This profiling technique is also capable of interpolation and extrapolation
beyond the phenotype information given in training data. As such, it can be used in many applications, including facilitating experimental design and detecting
confounding factors.

2) Phenotype association analysis problems are complicated by small sample size and high dimensionality. Consequently, phenotype-associated gene subsets
obtained from training data are very sensitive to selection of training samples, and the constructed sample phenotype classifiers tend to have poor
generalization properties. To eliminate these obstacles, we propose a novel approach that generates sequences of increasingly discriminative gene cluster
combinations. Our experiments on both simulated and real datasets show that the resulting cluster combinations are robust to perturbation in training samples
and have good sample phenotype classification performance.

3) Many complex phenotypes, such as cancer, are the product of not only gene expression, but also gene interaction. We propose an integrative approach to find
gene network modules that activate under different phenotype conditions and apply this approach to the study of cancer. Using our method on multiple cancer
gene expression datasets, we discovered cancer subtype-specific network modules, as well as the ways in which these modules coordinate. In particular, we
detected a breast-cancer specific tumor suppressor network module with a hub gene, PDGFRL, which may play an important role in this module.

\end{preface}



\chapter{Introduction}

\section{Integrative analysis of gene expression and phenotype association}
With the advancement of the high-throughput genotyping technologies, large amounts of genotype data have been accumulated in recent years. Such data include
gene expression, single nucleotide polymorphism, copy number variation, DNA mythelation, histone acetylation, alternative splicing, and proteomics. Since the
fundamental problem in morden genetics is the linking of genotype to phenotype. we are particularly interested in the associations between gene expression and
phenotype in this dissertation.

\section{Gene expression and phenotype data}
With the completion of the Human Genome Project, biology research is entering the post genome era. Although biologists have collected a vast amount of DNA
sequence data, the details that would explain how these sequences function still remain largely unknown. Genomes of even the simplest organisms are very
complex. Key questions still confront biologists, including 1) the functional roles of different genes and the cellular processes in which they participate; 2)
how genes are regulated, how genes and gene products interact, and how to identify interaction networks; and, finally, 3) how gene expression levels differ in
various cell types and states, as well as how gene expression is changed by various disease pathologies and their treatments. Biology used to be a data-poor
science. Nowadays, however, many of these questions can be addressed with the advent of more advanced techniques, such as gene expression.  Therefore,
investigators are now able to transform vast amounts of biological information into useful data.

This makes it possible to study gene function globally, and a new field, functional genomics, emerges. Specifically, functional genomics refers to the
development and application of global (genome-wide or system-wide) experimental approaches to assess gene function by making use of the information and
reagents provided by structural genomics.  Gene function may be divided into gene expression, which refers to the process of transcribing a gene's DNA sequence
into the RNA that serves as a template for protein production, and the level of gene expression, which indicates how active a gene is in certain tissues, at
certain times, or under certain experimental conditions.  Therefore, the monitoring of gene function can provide an overall picture of the genes being studied,
including their expression level and the activities of the corresponding protein under certain conditions. To accomplish this, functional genomics provides
techniques characterized by high-throughput or large-scale experimental methodologies combined with statistical and computational analysis of the results.

\begin{figure} [tph]
\includegraphics [width=1.0\columnwidth, viewport=1 1 800 536, clip] {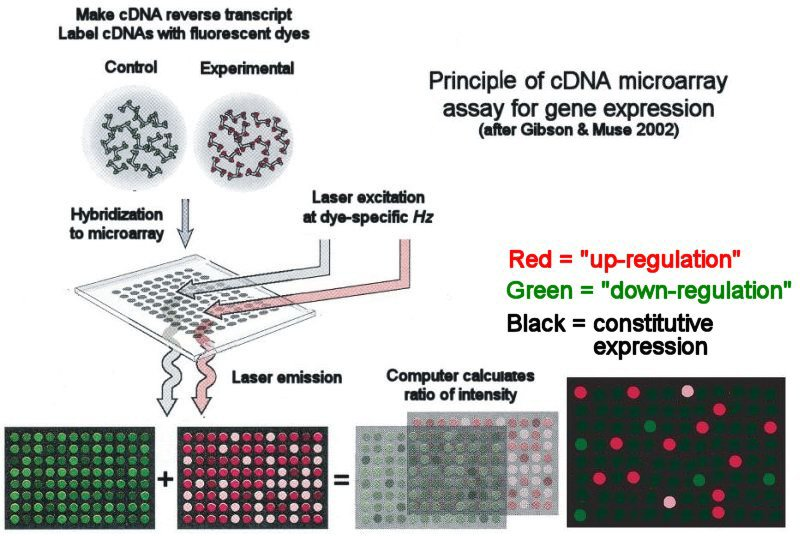}
\caption[Microarray technology]{A diagram illustrating how gene expression is obtained from microarray experiments. Figure obtained from http://www.mun.ca}
\label{fig:microarray_image}
\end{figure}

Among the several methods that have been developed to understand the behavior of genes, microarray technology is particularly important. It is used to
simultaneously monitor the expression levels of genes by several orders of magnitude, and many steps are involved in this technology. First, complementary DNA
(cDNA) molecules or oligos are printed onto slides as spots. Then, two kinds of dye-labeled samples, i.e., sample and control, are hybridized. Finally, the
hybridization is scanned and stored as images (Figure \ref{fig:microarray_image}). Using a suitable image processing algorithm, these images are then
quantified into a set of expression values representing the intensity of spots. Usually, the dye intensity may be biased by factors, such as physical property,
experimental variability in probe coupling and processing procedures, and scanner settings. To minimize the undesirable effects caused by this biased dye
intensity, normalization is done to balance dye intensities and make expression value comparable across experiments. Here the term comparable means that the
difference of any measured expression value of a gene between two experiments should reflect the difference of its true expression levels.

A phenotype is any observable characteristic of an organism. Phenotypes result from the activities of an organism's genes as well as the influence of
environmental factors and possible interactions between the two. Phenotypes are often described in natural languages in the literature. They can sometimes be
measured as continuous or categorical values. While it has thus far been very difficult to automatically extract phenotype information from the literature, the
Unified Medical Language System (UMLS) has been developed in recent years to provide facilities for natural language processing. As such, UMLS is a compendium
of many controlled vocabularies in the biomedical sciences. It provides a mapping structure among these vocabularies and thus gives researchers the ability to
translate among the various terminology systems. It is also a comprehensive thesaurus and ontology of biomedical concepts and consists of the following
components: 1) a Metathesaurus, the core database of the UMLS, which is a collection of concepts and terms from the various controlled vocabularies and their
relationships; 2) a Semantic Network which is a set of categories and relationships that are used to classify and relate the entries in the Metathesaurus; and
3) a SPECIALIST Lexicon which is a database of lexicographic information for use in natural language processing.  A number of supporting software tools are
also provided for this system.

\section[Integrative analysis of gene-phenotype association]{Integrative analysis of gene-phenotype \\ association}
In recent years, the accumulation of gene expression data has been a rapid process. For example,
the two largest public gene expression repositories, Gene Expression Ominibus (GEO) and Array
Express, already contain more than 500,000 samples, measuring genome wide gene expression levels of
various organisms under various phenotypic conditions.

Basically, these gene expression data repositories contain two types of information: gene expression and sample phenotype information. Expression information
based on microarrays are measured in continuous values. On the other hand, sample phenotype information is often described in natural language, or it can be
manually structured and recorded into a database. We develop methods of integrative analysis which combine these two types of information obtained from
different experimental sources, therefore enhancing confidence and providing a better understanding of the association between genotype and phenotype.

\section{Summary of our work}
As indicated above, we propose three approaches for the above analysis. 1) We propose the concept and a method of Automated Multi-dimensional Phenotypic
Profiling by using genotype information as a bridge. 2) We propose a method that can robustly search for combination of gene sets that provides true
discrimination and good prediction of sample phenotype classes. 3) We propose an integrative approach to characterize disease-specific pathways and their
coordination using gene coexpression network analysis. The first work primarily focus on phenotype prediction. Both the second and third work above involve
studying different aspects of gene phenotype associations. The second work focus on selecting phenotype associated combinations of gene sets, while the third
work particularly focus on genetic interaction mechanism that is associated to phenotype difference.

\subsection{Automated multi-dimensional phenotype profiling}
Phenotypes are complex and difficult to quantify in a high-throughput fashion. The lack of comprehensive phenotype data can prevent or distort
genotype-phenotype mapping. In Chapter \ref{Chp:PhenoProfileer} we describe \emph{PhenoProfiler}, a novel computational method that enables \emph{in silicon}
phenotype profiling. Drawing on the principle that similar gene expression patterns are likely to be associated with similar phenotype patterns,
\emph{PhenoProfiler} supplements the missing quantitative phenotype information for a given microarray dataset based on other well-characterized microarray
datasets.

We applied our method to 587 human microarray datasets covering more than 14,000 samples, and confirmed that the predicted phenotype profiles are highly
consistent with true phenotype descriptions. \emph{PhenoProfiler} offers several unique capabilities: (1) automated, multi-dimensional phenotype profiling,
facilitating the analysis and treatment design of complex diseases; (2) the extrapolation of phenotype profiles beyond provided classes; and (3) the detection
of confounding phenotype factors that would otherwise bias biological inferences. Finally, as no direct comparisons are made between gene expression values
from different datasets, the method can utilize the entire body of cross-platform microarray data. This work has produced a compendium of novel phenotype
profiles for the NCBI GEO datasets, which can facilitate an unbiased understanding of transcriptome-phenome mapping. By increasing the variaty and the quality
of phenotypes to be profiled, the continued accumulation of microarray data will further increase the power of \emph{PhenoProfiler}.

\subsection{Stable refinement of discriminative gene clusters}
Gene expression data are often used to identify the phenotype associated genes that are differentially expressed between samples of two phenotype classes. On
the other hand, since microarray datasets contain only a small number of samples and a large number of genes, it is usually desirable to identify small gene
subsets with distinct patterns between sample phenotype classes. Such gene subsets are highly discriminative in phenotype classification because of their
tightly coupling features. Unfortunately, gene subsets obtained from training data are often very sensitive to the selection of training samples, and the
classifiers constructed from such gene subsets usually tend to have poor generalization properties on the test samples as a result of the overfitting problem.

In Chapter \ref{Chp:diff_cluster}, we propose a novel approach \cite{xu2008sim} to correct these problems by combining supervised and unsupervised learning
techniques to generate increasingly discriminative gene cluster combinations in an iterative manner. Compared with existing approaches, our experiments on both
simulated and real datasets show that our method can produce a series of gene cluster combinations that are robust to training sample perturbation and have
good sample phenotype classification performance. This backward approach for refining a series of highly discriminative gene cluster combinations for the
purpose of sample phenotype classification has proven to be very consistent and stable when applied to various types of training data. In addition, these
combinations can be further used to study the underlying genetic interactions that lead to phenotype difference, and extended to integrate gene clusters
obtained from multiple datasets consisting of similar kinds of phenotype classification studies. Thus, the confidence that these clusters are truly
discriminative can be further enhanced.

\subsection{Disease-specific pathway identification}
The most common application of microarray technology in disease research is to identify genes differentially expressed in disease versus normal tissues.
However, in the case of complex diseases, it is well known that, phenotypes are determined by the underlying structure of genetic networks, not by individual
genes. Thus in many situations, it is the interaction of many genes that plays an important role in causing phenotypic variations. In Chapter
\ref{Chp:NetModule}, using cancer as an example, we develop a graph-based method ~\cite{xu2008iac} to integrate multiple microarray datasets to discover
disease-related co-expression network modules. We propose an unsupervised method that takes into account both co-expression dynamics and network topological
information to simultaneously infer network modules and phenotype conditions in which they are activated or de-activated.

Using our method, we have discovered network modules specific to cancer or subtypes of cancers. Many of these modules are consistent with or supported by their
functional annotations or their previously known involvement in cancer. In particular, we identified a module that is predominately activated in breast cancer
and is involved in tumor suppression. While individual components of this module have been suggested to be associated with tumor suppression, their coordinated
function has never been elucidated. Here by adopting a network perspective, we have identified their interrelationships and, particularly, a hub gene PDGFRL
that may play an important role in this tumor suppressor network. Consequently, by using a network-based approach, our method provides new insights into the
complex cellular mechanisms that characterize cancer and cancer subtypes. By incorporating co-expression dynamics information, our approach not only extracts
more functionally homogeneous modules than those based solely on network topology, but also reveals pathway coordination beyond co-expression.

\chapter{Automated Multi-dimensional Phenotypic Profiling} \label{Chp:PhenoProfileer}

\section{Introduction}

The fundamental aim of modern genetics is linking genotype to phenotype. With the rapid accumulation of genomics data, the lack of phenotype data has become the bottleneck of this process~\cite{freimer2003hpp}. Phenotyping, especially for human subjects, is a laborious process~\cite{lussier2007cap}. Moreover, researchers often gloss over the complexity of human phenotypes by reporting only those traits specifically relevant to their studies. For example, a given dataset may provide survival information but not the patients' ages.
Inferences derived from such data could be biased or even invalidated by undocumented or poorly documented phenotypic traits. Furthermore, most available phenotype characterizations are qualitative (categorical) rather than quantitative (continuous). This practice is problematic for two reasons: the boundaries between categories are often vague or arbitrary~\cite{oti2008pc}, and any phenotypic information distinguishing data within a category is lost.

In this paper, we address the above issues by developing \emph{PhenoProfiler}, a computational framework for predicting the quantitative phenotype information missing from a genomic dataset. In particular, this method associates each sample of a given dataset with the relative intensity of a specific phenotype trait. The quantitative measures of samples across the whole dataset is referred to as a \emph{phenotype profile} (PP). Examples include the body weights of individuals, degrees of malignancy in tumor samples, and
the quantitative responses of patients to drug treatments.

The principle of \emph{PhenoProfiler} is that similar genomic patterns are likely to be associated with similar phenotypic patterns~\cite{brunner2004sff}. Thus, we can supplement the (incomplete) phenotypic information in a given genomics dataset with traits recorded in other well-characterized datasets. In particular, we focus on the vast accumulated microarray data. The NCBI Gene Expression Omnibus (GEO)~\cite{barrett2007ngm}, for example, currently contains more than 2000 human microarray datasets that systematically document the transcriptome basis of phenotypes as diverse as heart diseases, mental illness, infectious diseases, and a variety of cancers.

\begin{figure} [tph]
\includegraphics [width=1.0\columnwidth, clip] {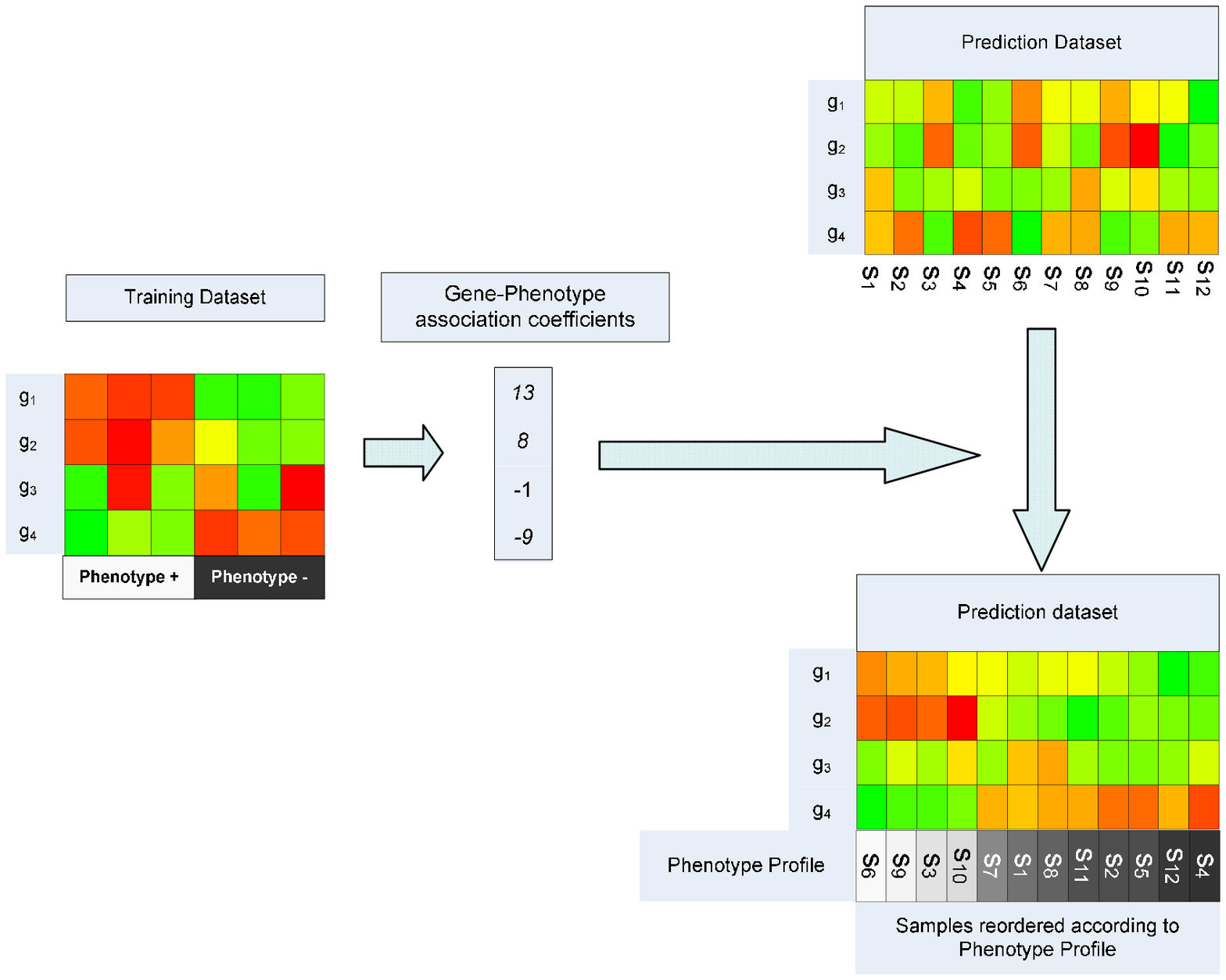}
\caption[The principles of \emph{PhenoProfiler}.]{The principles of \emph{PhenoProfiler}. Each row of an expression matrix corresponds to a gene, and each column corresponds to a
sample. The magnitude of gene expression is indicated by a color scale running from green (low) to red (high). From the training dataset (left), we obtain for each gene $i$ a
coefficient $w_i$ describing the degree of association between its expression level and the phenotype. Here, genes 1 and 2 have high positive coefficients. Gene 3 shows no clear
association with the phenotype, so its coefficient is close to zero. Gene 4 has a high negative coefficient. Therefore, genes 1,2 and 4 are signature genes of the phenotype. Given
a new dataset, for each sample we aim to estimate the relative intensity of the association between this sample and the phenotype such that the derived intensity values of all
sample are most similar to the expression values of signature genes. We term such a sample profile as a phenotype profile. The phenotype profile vector is depicted by the grey
cells, where brighter colors indicate estimated tighter association with the phenotype. After sorting the samples based on the phenotype profile, it can be seen that the
expression profiles of genes 1 and 2 are strongly correlated with phenotype profile, while that of gene 4 is anti-correlated.} \label{fig:method_intuition}
\end{figure}

The intuition behind our method is as follows. Given a training dataset with known sample description of a phenotype $P$, for each gene we can derive an association between its expression profile and this phenotype $P$. We denote those genes as \emph{signature genes}, the expressions of which are strongly associated with the phenotype in the training dataset. Given a new microarray dataset that is known to be related to the phenotype $P$, but the phenotype description of its individual samples are unknown, we aim to estimate
the PP by constructing a sample profile as a real-valued vector that is most similar to the expression profiles of those \emph{signature genes} in the new dataset. Figure~\ref{fig:method_intuition} illustrates this approach.

Since the information we borrow from the training dataset is only the association between the gene expressions and sample phenotypes, we do not directly compare the expression
values between the training and prediction datasets, thus bypassing the data incompatibility problem between cross-platform and cross-laboratory microarray datasets.
\emph{PhenoProfiler} can therefore use as many as possible microarray datasets in the public repositories in the training stage, and construct a new dataset's profiles for a wide
range of phenotypes.

We applied our method to 587 human microarray datasets, covering more than fourteen thousand microarray samples. The predicted phenotype profiles were highly consistent with known
phenotype descriptions. We showed that \emph{PhenoProfiler} can robustly provide multi-dimensional characterization of the phenotypes missing from a dataset, and can facilitate
the discovery of confounding factors for the transcriptome-phenotype mapping. The comprehensive phenotypic data generated by this method will vastly increase the value of
published and forthcoming genomics data.

\section{Results}

\subsection{Overview of method}

As illustrated in Figure~\ref{fig:method_intuition}, \emph{PhenoProfiler} consists of two steps. (1) Given a microarray dataset $D_1$ whose samples have known descriptions of the
phenotype $P$, for each gene $i$ we calculate a coefficient $w_i$ that indicates the degree of association between its expression profile and the phenotype $P$. The appropriate way to calculate $w_i$ depends on the structure of the phenotype descriptions. If the phenotype description is binary, i.e. the dataset compares two phenotype groups, we can simply use the two-sample $t$-statistic as $w_i$. If the phenotype description is continuous, we can use the correlation between the phenotype values and the gene expression values as $w_i$. Genes with a large magnitude of $w_i$ are termed the \emph{signature genes} of the phenotype $P$. (2) In order to predict the phenotype profile (PP) of a new microarray dataset $D_2$, we use the following constrained optimization approach. For a dataset with $m$ individual
samples, the PP is defined as a normalized, real-valued vector of $m$ values $(p_1, p_2, \ldots, p_m)^T$ (denoted $\mathbf{p}$). We also define the normalized expression vector of
the gene $i$ as $(e_{i1}, e_{i2}, \ldots, e_{im})^T$, denoted $\mathbf{e}_i$; and we denote the expression matrix containing all such expression vectors as $\mathbf{E}$. Given a
set of coefficients $w_i$ determined from the training data, the objective is to find a profile $\mathbf{p}$ that minimizes the weighted least-squares difference $\min \sum_i
|w_i| \sum_j (\mathrm{sgn}(w_i) e_{ij} - p_j)^2$. (The function $\mathrm{sgn}$ is $+1$ or $-1$ depending on the sign of its argument; we want the phenotype profile $\mathbf{p}$ to
be close to $\mathbf{e}_i$ when $w_i$ is positive, and close to $-\mathbf{e}_i$ when $w_i$ is negative.) A series of matrix computations (see the Methods section) yields the
optimal solution $\hat{\mathbf{p}}$, which is essentially the normalized weighted (by $w_i$) sum of gene expression values across samples.

To assess whether the predicted profile $\hat{\mathbf{p}}$ captures the expression trend of those signature genes, we calculate an association score $\hat{c}$, defined as the
Pearson's correlation between the coefficients $\mathbf{w}$ and $\mathbf{E}\hat{\mathbf{p}}$ (detailed explanation in Methods). To assess the statistical significance of
$\hat{\mathbf{p}}$, we compare the $\hat{c}$ to those calculated using the same expression matrix $\mathbf{E}$ and 1000 random permutations of coefficients $\mathbf{w}$.

\subsection{An illustrative case: reconstructing the temporal order of the yeast log-to-stationary growth transition}
As an illustrative example, consider the two microarray datasets GDS18 and GDS283. Both study the log-to-stationary growth transition of yeast, but with different microarray
platforms. Both datasets measure gene expression starting at the logarithmic phase and extending through the stationary phase. Here our goal is to predict changes in yeast
phenotype from log to stationary transition, responding to the depletion of nutrients. Naturally, the temporal order of the samples serves as a good means of validating the
prediction.

\begin{figure}[tph]
\centering
\includegraphics [width=\columnwidth] {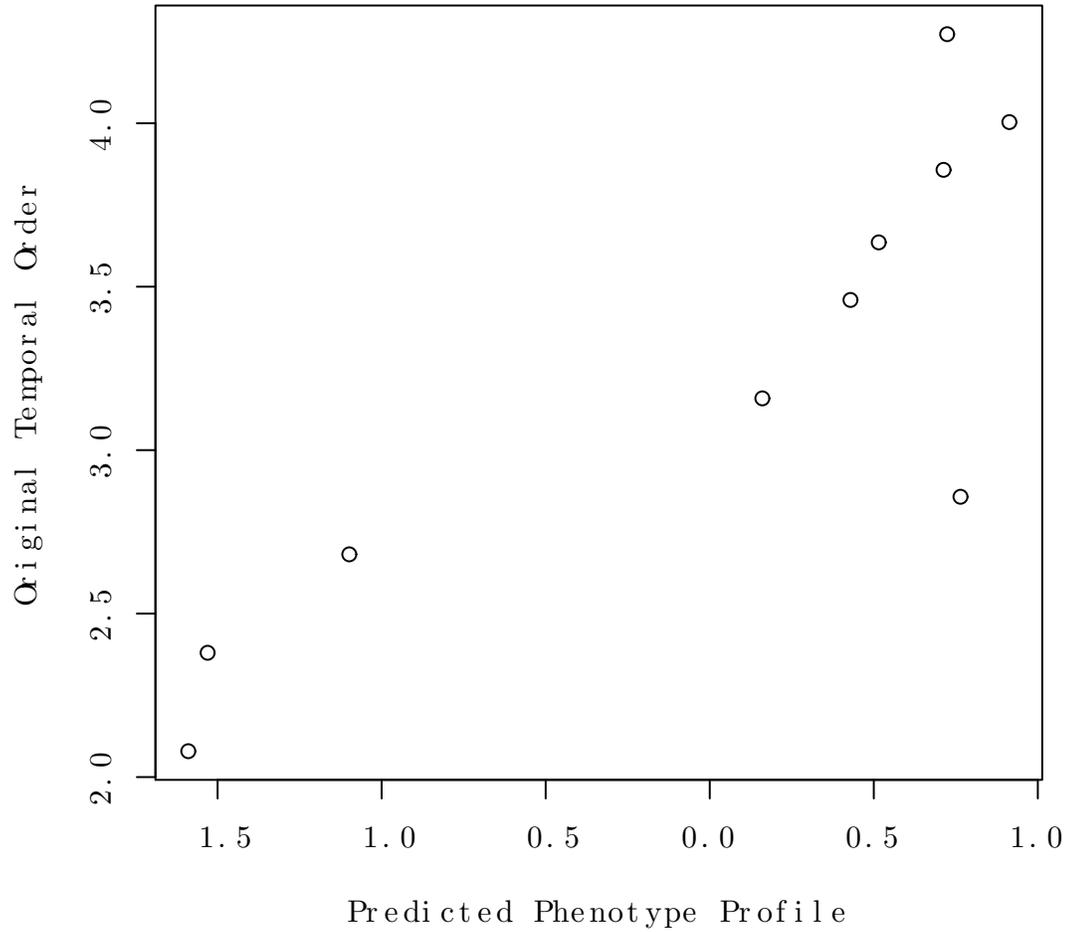}
\caption[Consistency between predicted and true phenotype profiles in GDS18.]{Consistency between predicted and true phenotype profiles in GDS18. Using GDS283 as the training
dataset, the predicted phenotype profile of GDS18 closely matches the original temporal order of the samples. The original temporal order is measured as the logarithm (base 10) of
minutes.} \label{fig:yeast_stationary_phase}
\end{figure}

Using one dataset for training, we computed the Spearman's rank correlation between individual gene expression profiles and the temporal order of the samples. These statistics are
used as the training coefficients. In the other dataset, we then predicted the phenotype response profile based on gene expressions, hoping to recover the correct temporal order
of samples. In both cases, the predicted PP was highly consistent with the actual sequence of samples (Spearman's rank correlation were 0.83 and 0.79, depending on which dataset
was used for training). Figure~\ref{fig:yeast_stationary_phase} shows the predicted and the original sample order of dataset GDS18. Two subgroups are visible in the predicted
profile, accurately reflecting the logarithmic and stationary phases. The sole exception is at the transition between the two phases.

Intriguingly, experiment GDS283 stopped taking measurements only 17 hours after the yeast entered the stationary phase. Experiment GDS18, on the other hand, continued measurements
for another 12 days. It is remarkable that a phenotype signature derived from GDS283 can accurately sort the phenotype progression of GDS18. This result demonstrates that the
essential physiological changes occurring within and between the logarithmic and stationary growth phases can be extrapolated as well as interpolated.

\subsection{Large-scale prediction of phenotype profiles}

To test the general applicability of our method, we performed a large-scale analysis of 587 human microarray datasets (see the Methods section for details on data collection and
processing). Datasets containing at least two disjoint sample groups, representing a phenotype and its baseline ($P$ and $P'$), each with at least 10 samples, were selected as
training datasets ($D_1$). If a dataset contains $n$ sample groups, we can generate $n \choose 2$ distinct training datasets. Since the phenotype values in these datasets are
categorical, we use the two-sample $t$-statistic as coefficients $\mathbf{w}$. By setting the threshold $p$-value for the predicted PP to 0.001 and the association score $\hat{c}
\geq 0.25$, a total of 37,852 novel PPs were associated to the 587 datasets.

To validate the method, for each training dataset $D_1$ we also need a testing dataset $D_2$ that contains the sample descriptions on exactly the same phenotypes $P$ and $P'$.
Among all 587 datasets, we only identified 4 training-testing dataset pairs meeting this criterion, in which each of the testing datasets also contains two sample groups of $P$
and $P'$. To assess whether the predicted phenotype profile is consistent with the known distribution of phenotypes $P$ and $P'$ in the testing dataset, we used the
\emph{Wilcoxon} rank sum test. Specifically, in the test dataset, the two sample groups' ($P$ and $P'$) predicted phenotype values are compared using \emph{Wilcoxon} rank sum
test. A small Wilcoxon $p$-value indicates that there is a significant difference between the distributions of predicted phenotype values for the two groups, therefore the
predicted profile is consistent with known phenotype information. Among the 4 training-testing pairs, all predicted PPs were highly consistent with the known phenotype groups
(Wilcoxon $p$-value $ < 10^{-4}$).

In order to obtain a general assessment using more validation data, we relaxed the requirement that the description of the testing dataset exactly matches the training data
phenotypes. In fact, if the phenotypes of a given dataset were even moderately similar to the training phenotype, the predicted profiles were found to agree well with known
phenotype groups in the testing dataset. This implies a strong interdepedence among related phenotypes. We quantify the similarity between the training and testing phenotype with
two measures: (1) the percentage $\gamma$ of \emph{Unified Medical Language System} (UMLS) concepts of the merged sample group descriptions of $D_1$ shared with the dataset
description of $D_2$; and (2) the similarity between the descriptions of corresponding sample groups in $D_1$ and $D_2$, denoted as $s$. The latter is defined as the cosine of the
angle between two \emph{term frequency-inverse document frequency (tf-idf)} vectors of mapped UMLS terms (See the Methods section for details). Following these measurements, we
identified 32 training-testing dataset pairs with similarity thresholds $s
> 0.4$ and $\gamma > 0.6$.  Among these, 81\% of predicted phenotype
profiles were consistent with prior phenotype descriptions (Wilcoxion test $p$-value $ < 0.05$). This result highlights the effectiveness of our method in exploiting the
interdependence of similar phenotypes.

We further studied the robustness of our method against the size of the training dataset. We randomly selected a training dataset and a test dataset, and then calculated the
correlation between the PP constructed with the original training dataset and that with certain amount of training samples randomly removed. Repeating this test 10,000 times with
10\% (and 20\%) sample removal produced an average correlation of 0.98 (and 0.95) between the resulting PPs and those without any samples removed. Even for those training datasets
with a small size of 10 samples in each of the two phenotype groups, the obtained PP correlations were still greater than 0.9 for both 10\% and 20\% sample removal, demonstrating
the robustness of our method.

\subsection{Multi-dimensional profiling of complex phenotypes}
As previously mentioned, a total of 37,852 PPs were derived and assigned to the 587 datasets. On average, each dataset is assigned 65 PPs. In some cases, related training datasets
generated highly correlated PPs, further enhancing our confidence in the prediction. Two examples are described below.

Dataset GDS2855 studies various forms of muscular dystrophy. Three training sets (GDS609, 610, and 612) generated highly correlated PPs (average correlation 0.88) for GDS2855. All
three training datasets describe the difference between Duchenne muscular dystrophy and normal muscle tissues, although they were measured with different platform technologies.
Furthermore, all 3 predicted PPs were highly consistent with the original sample description of GDS2855 (Wilcoxion test $p$-value $ < 10^{-6}$).

Dataset GDS1962 studies gliomas of different grades, and was assigned four highly correlated PPs (average correlation 0.9) by datasets GDS1975, 1976, 1815, and 1816. All four
training datasets focused on comparing grade III and grade IV glioma samples. Remarkably, the predicted PPs not only did a good job of separating grade III from grade IV samples
in the testing dataset, but also separated grade II from grade III samples. In addition, the separations followed the order of tumor grades. This example shows that our method
captures the essential difference between high- and low-grade tumors, and thus can be extrapolated to tumors of grades beyond those represented in the training data. This ability
to extrapolate from the training dataset represents a significant advantage over traditional classification methods.

\begin{figure}[tph]
\centering
\includegraphics [width=\columnwidth] {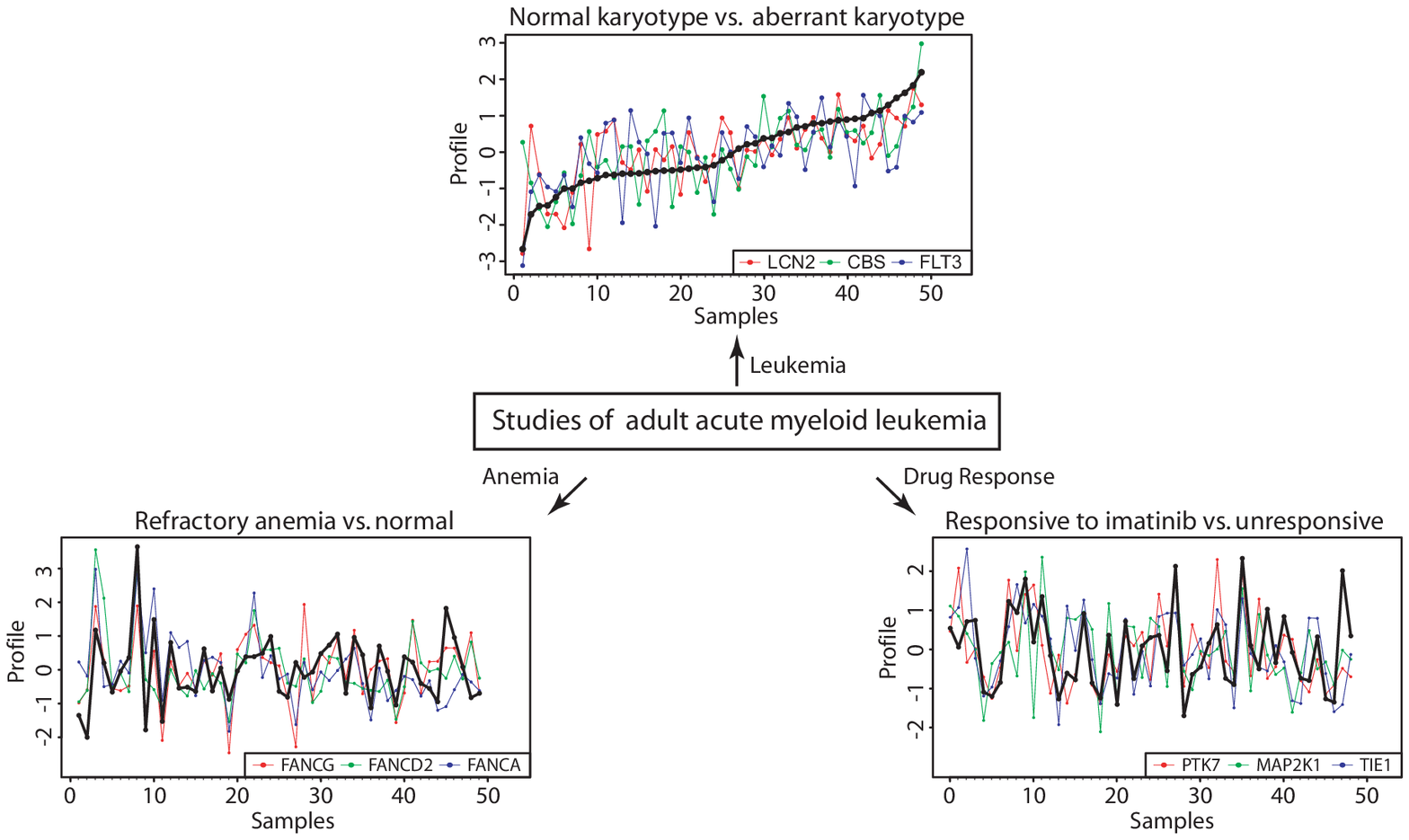}
\caption[Multi-dimensional profiling of the dataset GDS843.]{Multi-dimensional profiling of the dataset GDS843 that compares patients of adult AML with abnormal karyotypes to
those with normal karyotypes. Three different training datasets produced mutually uncorrelated phenotype profiles (black curves) that could be assigned to GDS843. The three most
significantly correlated expression profiles of genes known to be related to the respectively profiled phenotypes are superposed in the three primary colors. For clarity, the
samples are ordered according to the first PP, and the expression profiles of negatively correlated expression profiles have been reversed. } \label{fig:multi_dim_profiling}
\end{figure}

A testing dataset is often (78\% of cases) assigned multiple uncorrelated PPs (correlation $<$ 0.1) describing different properties of a complex phenotype. For example, dataset
GDS843 contains 49 samples comparing patients with abnormal karyotypes to patients with normal karyotypes to study adult acute myeloid leukemia (AML). The samples were collected
from peripheral blood or bone marrow. Its predicted phenotypes include three uncorrelated profiles (see Figure ~\ref{fig:multi_dim_profiling}), which are detailed below.

\begin{enumerate}
\item Training dataset GDS842 also studied abnormal versus normal
karyotypes in adult AML patients. The derived phenotype profile is consistent
with known sample description of this phenotype in GDS843 (Wilcoxon $p$-value
0.04), thus validating our method.

\item Training dataset GDS2118 compared individuals with refractory anemia
to normal individuals. The PP trained by this comparison is highly correlated (correlation $>$ 0.9) with two other PPs that also come from training datasets that studied
refractory anemia. In fact, the recently proposed WHO classification of hematologic malignancies merged the disease refractory anemia with excess blasts in transformation (RAEB-T)
into the category AML. However, this new disease classification is controversial. Although RAEB-T and AML share similar clinical parameters, a study pointed out that their
biological bases are different (e.g. RAEB-T is distinguished from AML by a significant increase in apoptosis), and it suggested that RAEB-T should be regarded as a distinct
disease entity ~\cite{albitar2000dbr}. Therefore, the derived PP may uncover hidden patient information and possibly help to differentiate RAEB-T from AML, which could further
lead to improved treatment design.

\item Training set GDS1221 studies the patient response to the drug
Imatinib. Imatinib was designed to treat chronic myeloid leukemia by reducing the tyrosine kinase activity of the well-known bcr-abl fusion gene. Our phenotype profile could
therefore be used to identify patients that would be more likely to respond to Imatinib treatment.

\end{enumerate}

In summary, while the first PP serves as an internal validation of the method, the other two PPs provide novel insights into the pathologic and therapeutic properties of sample
phenotypes in the dataset GDS843. The specific phenotype properties represented by the above PPs can be further confirmed by examining genes whose expressions are significantly
correlated ($p$-value $<$ 0.001) with these profiles. For example, the AML PP (number 1 above), has 10 significantly correlated genes that are known to be associated with the UMLS
concept ``Leukemia, Myelocytic, Acute.'' A particularly interesting gene is FLT3. A study suggested that in patients with karyotype alterations, a recipricol translocation was not
sufficient to cause acute promyelocytic leukemia, and that an additional mutation in FTL3 may be required ~\cite{delourdeschauffaille2008apl}. For the refractory anemia activation
profile, there are 12 significantly correlated genes known to be involved in ``Anemia.'' These include three Fanconi Anemia genes (FANCA, FANCD2, FANCG) as well as TGFB1, which
may affect the progression of refractory anemia specifically ~\cite{balog2005cit}. Among the genes correlated with the Imatinib response profile, three are tyrosine kinases, which
is consistent with the target of Imatinib ~\cite{deininger2003stt}. This example demonstrates the power and specificity of our multi-dimensional phenotype profiling approach by
utilizing the large number of microarray datasets.

\subsection{Discovery of hidden confounding factors in microarray studies}

Due to the scarcity of phenotype information in many microarray datasets, confounding phenotype variables may not be well documented. Thus, caution should be exercised in deriving
inferences from microarray datasets. The following cases provide representative scenarios.

Dataset GDS1673 examines normal lung tissue from 23 donors, including smoking and non-smoking individuals. Interestingly, we found that a predicted PP trained on male vs. female
skeletal muscle samples (dataset GDS914) was able to separate the smoking and non-smoking samples of GDS1673 (Wilcoxon $p$-value 0.0002). After obtaining additional phenotype
information on the GDS1673 subjects, it turns out that among non-smokers, which made up almost two-thirds of the sample, females outnumbered males by more than two to one, while
among smokers the numbers of the two genders were approximately equal. Thus, simply comparing the expression profiles of the smoking versus non-smoking groups would not derive the
signature of smoking, but rather the mixed signatures of smoking and gender.

As another example, the goal of the GDS1887 study was to build a prognosis model for rectal cancer cells responding to radio therapy. According to the GEO annotation, its 46
samples had been separated into \emph{training} and \emph{test} groups for model construction and validation. Surprisingly, we found that the original \emph{training} and
\emph{test} samples from this study could be well separated (average Wilcoxon $p$-value 0.001) by four highly correlated PPs (average correlation 0.95). All of those four PPs come
from training datasets that compare the cancer to normal tissue or that compare cancers of different malignancies. This strongly suggest that there were systematic differences in
cancer malignancy between the training and testing samples, even though they were supposed to be generated by random partition. Any such sampling bias would negatively impact the
accuracy of the prognosis model.

Of course, sampling bias can often be traced to the very limited availability of phenotype data in the first place. Our compendium of predicted phenotype profiles \\
(http://zhoulab.usc.edu/PhenoProfiler) provides a comprehensive description of a large proportion of the datasets in the GEO database. This knowledge can facilitate an unbiased understanding of the transcriptome-phenome mapping. It can also serve as the starting point for the identification of molecular mechanisms shared by different diseases and
phenotypes.

\section{Discussion}
Phenotypes are complex, and difficult to quantify in a high-throughput fashion. The lack of comprehensive phenotype data can prevent or distort genotype-phenotype mapping. This
paper describes a novel approach to perform \emph{in silicon} phenotype profiling. Our method provides numerous advantages which we outline here. (1) For most datasets we were
able to predict multiple phenotype profiles, which could help researchers to reveal different aspects of complex diseases and facilitate treatment design. (2) We can provide a
quantitative phenotype description of the sample characteristics. Although ``categorical'' phenotype description is prevalent, in reality phenotypes constitute a continuous
spectrum. (3)  Our method can extrapolate the profiling to classes beyond those represented in the training data, as illustrated in the glioma case study. This is an advantage
over traditional classification methods. (4) \emph{PhenoProfiler} avoids direct comparison of gene expression values from different datasets, and thus can utilize \emph{almost
all} available microarray data regardless of platform or laboratory. In contrast, traditional regression methods cannot be directly applied to microarray datasets from different
platforms.

The continued accumulation of microarray data will further increase the power of \emph{PhenoProfiler} in two aspects: the variety of phenotypes to be profiled, and the confidence
of its predictions. The latter benefit derives from having several mutually correlated PPs from similar datasets. The principles of our method can be easily applied to other types
of genomics data (e.g. proteomics or metabolomics) as they become increasingly available. The present work focuses on linear gene-phenotype associations, but more complex
relationships can be devised depending on the data characteristics.

Our univariate method for selecting the weights from the training sample is only one of many possible approaches. For example, one could consider constructing weights using a
multivariate procedure that takes into account correlations among the gene expression levels, such as Fisher's linear discriminant procedure (i.e Discriminant Function Analysis
for two groups). However, such an approach requires estimating a covariance matrix for the gene expressions which is not practical given that there are thousands of genes and a
limited number of samples (typically on the order of 10) per dataset. In fact Fan and Fan ~\cite{fan2008hdc} prove that, when the dimension of the feature space is high, a
univariate two-sample t-test procedure, similar to our approach, is often superior to a multivariate method. Alternatively, when the important phenotype information can be
characterized using a small number of linear combinations of the genes, dimension reduction techniques like Nonnegative Matrix Factorization ~\cite{brunet2004mam, tamayo2007mpc}
may also produce meaningful phenotype predictions.

\section{Methods}

\subsection[Predicting phenotype profiles by constrained optimization]{Predicting phenotype profiles by constrained \\ optimization}

 From a training microarray dataset, we derive a vector $\mathbf{w}=(w_1,w_2,\ldots,w_n)^T$ that contains the gene-phenotype association coefficients of $n$ genes. Given a new dataset with $n$ genes and $m$ samples, and the
normalized gene expression matrix $\mathbf{E}=(e_{ij})_{n\times m}$, we aim to obtain the optimal \emph{Phenotype Profile (PP)} of the $m$ samples, where PP is a normalized,
real-valued vector $\mathbf{p}=(p_1,p_2,\ldots,p_m)^T$ that show high similarity to the expression profiles of those genes that have high magnitude of gene-phenotype association
coefficients $w$ (termed \emph{signature genes})

\begin{eqnarray}
Q_1: & \underset{\mathbf{p}\in\mathbb{R}^m}{\min} & d(\mathbf{E}, \mathbf{w}, \mathbf{p}) = \sum_i
|w_i| \sum_j (\mathrm{sgn}(w_i) e_{ij} - p_j)^2
\nonumber\label{Eqn:minimization1}\\
&\text{subject to} & \sum_j p_j = 0 \nonumber\label{Eqn:meanzero1}\\
& & \frac{1}{m-1} \sum_j{p_j^2} = 1 \nonumber\label{Eqn:varunit1}
\end{eqnarray}

Let $\mathbf{b} = (b_1, b_2, \ldots, b_m)^T$ be the weighted sum of gene expression values for each sample, $b_j = \sum_i w_i e_{ij} $. The following theorem provides the solution
to the minimization problem.
\\
\\
\noindent \textbf{Theorem:} The solution $\hat{\mathbf{p}}$ to problem $Q_1$ is a vector that is the normalized form of $\mathbf{b}$. That is,
\begin{equation}
\hat{\mathbf{p}} = \frac{\mathbf{b}-\overline{\mathbf{b}}}{\sigma(\mathbf{b})} \nonumber\label{Eqn:solution}
\end{equation}
\noindent where $\overline{\mathbf{b}}$ and $\sigma(\mathbf{b})$ are the mean and standard deviation of $\mathbf{b}$ respectively.

\textbf{Proof:} By expanding the function $d$, we have 

\begin{eqnarray}
d(\mathbf{E}, \mathbf{w}, \mathbf{p}) = \sum_i |w_i| \sum_j e_{ij}^2 + \sum_i |w_i| \sum_j p_j^2 - 2 \sum_i w_i \sum_j
e_{ij} p_j \nonumber
\end{eqnarray}

\noindent Since $\mathbf{p}$ is normalized and $\mathbf{E}$ and $\mathbf{w}$ are fixed, the first two terms are fixed. So the minimization problem $Q_1$ can be simplified to an
equivalent maximization problem:

\begin{eqnarray}
Q_2: & \underset{\mathbf{p}\in\mathbb{R}^m}{\max} & c(\mathbf{E},\mathbf{w},\mathbf{p}) = \mathbf{w}^T \mathbf{E} \mathbf{p} \nonumber\label{Eqn:maximization1}\\
&\text{subject to} & \mathbf{p}^T\mathbf{1}=0 \nonumber\label{Eqn:meanzero2}\\
& & \mathbf{p}^T\mathbf{p}=m-1 \nonumber\label{Eqn:varunit2}
\end{eqnarray}

\noindent where $\mathbf{1}$ is the vector whose elements are all 1.

Let $\mathbf{b} = \mathbf{E}^T \mathbf{w}$. Let the Lagrangian function for $Q_2$ be

\begin{equation}
L(\mathbf{p},\lambda_1,\lambda_2) = \mathbf{b}^T\mathbf{p} + \lambda_1\mathbf{p}^T\mathbf{1} + \lambda_2(\mathbf{p}^T\mathbf{p}-(m-1))\nonumber \label{Eqn:Lagrangian}
\end{equation}

\noindent where $\lambda_1$ and $\lambda_2$ are Lagrangian multipliers. According to the Karush-Kuhn-Tuker conditions~\cite{bazaraa1993} (as the functions
$\mathbf{b}^T\mathbf{p}$, $\mathbf{p}^T\mathbf{1}$, and $\mathbf{p}^T\mathbf{p}$ are all convex), the solution to Equation~\ref{Eqn:Gradient-Eqns} contains the global optimum of
$Q_2$,

\begin{equation}
\nabla L(\mathbf{p},\lambda_1,\lambda_2) = 0 \label{Eqn:Gradient-Eqns}
\end{equation}

Equation~\ref{Eqn:Gradient-Eqns} results in two solutions: $\mathbf{p}=\pm\frac{\mathbf{b}-\overline{\mathbf{b}}}{\sigma(\mathbf{b})}$. Since $Q_2$ is a maximization problem, it
is easy to show that the solution of $Q_2$ is $\hat{\mathbf{p}}=\frac{\mathbf{b}-\overline{\mathbf{b}}}{\sigma(\mathbf{b})}$, and so is the solution of $Q_1$.

$\hat{\mathbf{p}}$ is regarded as the PP of the new dataset because among all vectors in $\mathbb{R}^{n}$, $\hat{\mathbf{p}}$ is the one that most resembles the normalized
expression profiles of the signature genes that were defined by the training data.

We calculate an association score $\hat{c}$ as the Pearson correlation between $\mathbf{w}$ and $\mathbf{E} \hat{\mathbf{p}}$. The score is derived from the maximization problem
$Q_2$. $\mathbf{E} \hat{\mathbf{p}}$ provides the association between expression profiles and the predicted phenotype profile in the testing dataset. Thus, higher $\hat{c}$
indicates higher consistency of gene-phenotype associations derived from the training and testing datasets.

\subsection {Data collection and processing}
We collected 587 human microarray datasets, each containing at least 5 samples, from the NCBI Gene Expression Omnibus (GEO) ~\cite{barrett2007ngm}. For data generated with the
Affymetrix platforms, we increased any values less than 10 to 10 and performed a log transform of the gene expression values. For genes with multiple probesets present, the
expression values of those probesets were averaged. We then normalized each dataset by converting the expression values of each gene to Z-scores (zero mean and unit variance). The
587 datasets yielded 537 training datasets. A training dataset contains two disjoint sample groups, each of which contains at least 10 samples, representing a phenotype and its
baseline. When performing PP prediction, we discard those training-testing dataset pairs that share fewer than 100 genes in common.

\subsection{Automatic processing of phenotype annotations}
In GEO, a dataset is usually annotated by a short description paragraph; a sample group is annotated by a word or a short phrase; and a sample is usually annotated by one
sentence. To systematically categorize the phenotype information associated with each microarray dataset, we used the Unified Medical Language System
(UMLS)~\cite{bodenreider32uml, butte2006cai}. We mapped the dataset description, sample group descriptions, and sample descriptions onto UMLS concepts via the MetaMap Transfer
program~\cite{aronson2001emb}. To reduce noise we focused on disease-related concepts, including the MeSH vocabulary and the semantic types ``Pathologic Function'', ``Injury or
Poisoning'', ``Anatomical Abnormality'', ``Body Part, Organ, or Organ Component'', ``Tissue'', and ``Cell''. In general, the higher a concept is on the UMLS hierarchy, the broader
is the concept. Disease concepts at the fine granularity level may be associated with more clinical significance. In order to infer higher-order links between datasets, all of the
ancestor concepts of mapped concepts were included.

\subsection{Measuring phenotype annotation similarity}

We measure phenotype similarity between sample groups (the two groups of a training dataset and the two groups of a testing dataset) by the following procedure. 1) For each group,
we map its title and member sample descriptions onto UMLS concepts. 2) We then construct a term frequency-inverse document frequency (tf-idf) vector~\cite{salton1988twa} for each
sample group. 3) Suppose that $\mathbf{U}_{11}$ and $\mathbf{U}_{12}$ are two tf-idf vectors corresponding to the two sample groups in the training dataset, and $\mathbf{U}_{21}$
and $\mathbf{U}_{22}$ are tf-idf vectors corresponding to the two sample groups in the testing dataset. The similarity between the sample groups is then calculated as
$\max(<\mathbf{U}_{11}, \mathbf{U}_{21}> + <\mathbf{U}_{12}, \mathbf{U}_{22}>, <\mathbf{U}_{11}, \mathbf{U}_{22}> + <\mathbf{U}_{12}, \mathbf{U}_{21}>)$, where $<\mathbf{a},
\mathbf{b}>$ denotes the cosine similarity~\cite{lage2007hpi} of two vectors $\mathbf{a}$ and $\mathbf{b}$, calculated as a normalized dot product $<\mathbf{a}, \mathbf{b}> =
\frac{\mathbf{a} \cdot \mathbf{b}} {\| \mathbf{a} \| \| \mathbf{b} \|} $. Essentially, this measure identifies the best match between the sample groups in the training dataset and
testing dataset while taking into account the possibility that the groups could be matched in reverse order.

\subsection{Calculation of TF-IDF vectors of sample groups}

Note that a sample group is called ``subset'' in NCBI GEO. Annotation of a sample group includes UMLS concepts, mapped from both (1) its text title and (2) all
its samples' text description. In our experiment, each sample group is viewed as document and each UMLS concept is viewed as a term. Thus, given 587 human
microarray datasets, we collected 4065 documents (sample groups) and 1277 terms (UMLS concepts).

In NCBI GEO annotation of sample groups, its title and each samples' description share almost the same set of UMLS concepts. Our practical experiments showed
that, in such complicated annotation environment, the frequency of a UMLS concepts in a sample group does not indicate the importance of the UMLS concept. So
we used only 0 or 1 to represent the term frequency of each UMLS concept. Then IDF of each term (UMLS concept) is counted based on the binary TF vector of each
document (sample group). Details of how to get TF-IDF is presented in the following formulas:

\begin{align}
&TF_{\text{term, document}} = & & \left\{\begin{array}{ll} 1, & \text{if this term occurs in the document;}\\ 0, & \text{otherwise.}\end{array}\right.\\
&IDF_{\text{term}} = & & \log\left(\frac{\text{number of all documents}}{\text{number of documents this term occurs}}\right) \\
&TFIDF_{\text{term, document}} = & & TF_{\text{term, document}}\times
IDF_{\text{term}}
\end{align}

In practical experiments, we noticed that a large portion of UMLS concepts are shared by two sample groups in the same dataset (e.g., $U_{11}$ and $U_{12}$
from the training dataset). Ideally, we expect that annotation of two sample groups from the same dataset should reflect their phenotype differences, not
commonness. Thus, we mask by zero in TF-IDF vectors, the common UMLS concepts shared by the same-dataset sample groups, e.g, $U_{11}$ and $U_{12}$ (or $U_{21}$
and $U_{22}$ from testing dataset), when computing annotation similairty.

\chapter{A Stable Iterative Method for Refining Discriminative Gene Clusters} \label{Chp:diff_cluster}
\section{Introduction}
Microarray has become an important tool for identifying genes that discriminate sample phenotype classes because of its power of monitoring the expression
levels of thousands of genes in a single experiment. Finding discriminative genes with gene expression data is actually the feature selection problem in
classification theory. From the machine learning point of view, it is critical since the gene expression datasets usually contain a small number of experiments
(called samples) and a large number of genes (called features) in each experiment. The selected highly discriminative genes after filtering out those
non-representative genes which may dilute the pattern in classification computation can be further studied for the investigation on the biological mechanisms
that are responsible for sample phenotype class distinction.

A number of efforts have been put in searching effective gene selection methods (For example \cite{golub1999mcc, guyon2002gsc, xiong2001bif, ambroise2002sbg,
furlanello2003ebg, zhang2006rsf}). Due to the small-sample size and high-dimension properties of the sample phenotype classification problem, it is not
difficult to find a feature subset that can perfectly discriminate all the samples \cite{xu2003gsf}. In fact, theoretical study in \cite{cover1965gas} showed
that even for the non-informative, randomly generated dataset, the expected size of a feature subset that can linearly discriminate all the $n$ samples is just
$(n+1)/2$. In microarray data analysis, there can be a large number of highly discriminative subsets containing only a couple of genes; and each individual
gene in such a subset is not necessarily highly discriminative. For example, we observed  by exhaustive search  that there are as many as 10,173 perfect 3-gene
subsets for sample phenotype classification with the weighted voting method proposed by Golub et al and with their proposed training-test split
\cite{golub1999mcc}; and these gene subsets cover 3,337 genes (93.4\% of all the 3,571 genes in the datasets after preprocessing). This observation suggests
that a method of finding a highly discriminative compact gene subset is not enough. The variability of the subsets found by such a method likely hinders the
discovery of real interaction among the genes given that the method is usually sensitive to both the choice of samples and noise in the microarray data.

The fundamental limit and challenges mentioned above motivates us to design more robust methods by taking into account the expression similarity information
among genes. In this paper, we identify a series of discriminative gene clusters by running clustering and feature selection processes iteratively, where the
centroids of the clusters are used to form predictors. This work also shows that the predictor constructed in this way is more stable and less sensitive to the
choice of training samples. Because biological functions are usually resulted from collective behavior and coordinated expression of a group of genes rather
than that of an individual gene, genes grouped according to their co-expression pattern may be more powerful in revealing gene regulation mechanisms.

Our approach to generate discriminative gene clusters is a combination of supervised and unsupervised technique  In recent years, Jornsten and Yu
\cite{jornsten2003sgc} and Dettling and Buhlmann \cite{dettling2002scg} proposed similar combination approaches. However, there are major differences between
their methods and our method.  We use a multivariate approach for cluster selection, while Dettling and Buhlmann \cite{dettling2002scg} employed a univariate
approach, which assumes the independence of the contribution of clusters to sample phenotype classification. Although such hypothesis reduces computational
complexity for large datasets, the accuracy is compromised since the complex biological interaction among gene clusters is not properly reflected. We exploit a
multivariate approach in the content of gene expression analysis since it accounts for the joint contribution of clusters to sample phenotype classification.
It is known that such complex phenotype like cancer is resulted from not expression of individual genes but rather the underlying genetic networks. Thus a
multivariate approach that emphasize the combinatorial effect of variables is more suitable to the exploration of the interaction among variables. Our method
differs from \cite{jornsten2003sgc} in the following two aspects: Although both works adopt multivariate approach, first of all, in their information-based
approach, clustering and cluster selection are done simultaneously, resulting in a set of clusters optimizing the Minimum Description Length. In comparison,
our computation-oriented approach is a refining process where clustering and cluster selection are performed alternatively in each iteration step with better
and better results. Secondly, the clusters generated with Jornsten and Yu's approach include both active and inactive ones. Here, active clusters are those
whose centroids are relevant to sample phenotype classification, and inactive ones are not. Our method is essentially a backward approach
\cite{ambroise2002sbg}. It iteratively eliminates the less active clusters and re-clusters the remaining genes in the active clusters, reducing the negative
influence of non-discriminative clusters on the sample phenotype classification.

Our program outputs a series of cluster sets that have increasing discrimination power for training samples without losing prediction power on the test
samples, as indicated in our experimental results. It achieved similar or better prediction accuracy than the known methods aforementioned for most of the
tested datasets in our validation process. More importantly, our test shows that the centroids of the output clusters using different sets of training samples
are stable and consistently achieve significant proximity to the global optimal gene clusters obtained by using all the samples. Another advantage of our
method is that it provides researchers with flexibility to decide which cluster set should be chosen for their purpose.

\section{Results}
We implemented the algorithm (described in the Methods section) as MATLAB functions. It runs on a PC with the Windows operating system. The SVM program written by Cawley was
downloaded from the website http://theoval.sys.uea.ac.uk/~gcc/svm/toolbox. In this section, we present the detailed test results on both simulated and Leukemia AML/ALL datasets
\cite{golub1999mcc}. We also have tested our method on other real datasets and compared the performance of our algorithm with those reported in the previous literature. The
details of the performance measures are described in the Method section.

\subsection{Simulated datasets}
We generated 100 simulated binary classification datasets using a simple stochastic model. Each simulated dataset contains 100 samples evenly split into two
sample phenotype classes. Both training and test samples contain 25 samples in each class.

Each dataset contains of 400 genes evenly divided into four gene clusters. Two of the four clusters are relevant to sample phenotype classification and these
two discriminative clusters $C_1$ and $C_2$ contribute to sample phenotype classification independently. Their centroids $\chi(C_1)$ and $\chi(C_2)$  are
generated according to the sample phenotype class labels. Each component of $\chi(C_1)$ in a position is generated according to normal distributions N(1, 0.5)
or N(-1, 0.5) depending on whether the corresponding sample is in class 1 or class -1, while each component of $\chi(C_2)$  generated according to N(-1, 0.5)
if the sample is in class 1 and N(1, 0.5) otherwise. Similarly, the centroids of the non-discriminative clusters $C_3$ and $C_4$ are generated according to the
normal distribution N(1, 1) and N(-1,1) regardless of the samples' class. For each $i$=1, 2, 3, 4, the expression values of a gene in the cluster $C_i$ are
generated according to the multivariate normal distribution $N(\chi(C_i), d_i / 4)$, where $d_i = \min_{j\neq i}d(\chi(C_i), \chi(c_j))$.

\begin{figure}[tph]
\centering
\includegraphics [width=\columnwidth, viewport=1 1 1200 1072, clip] {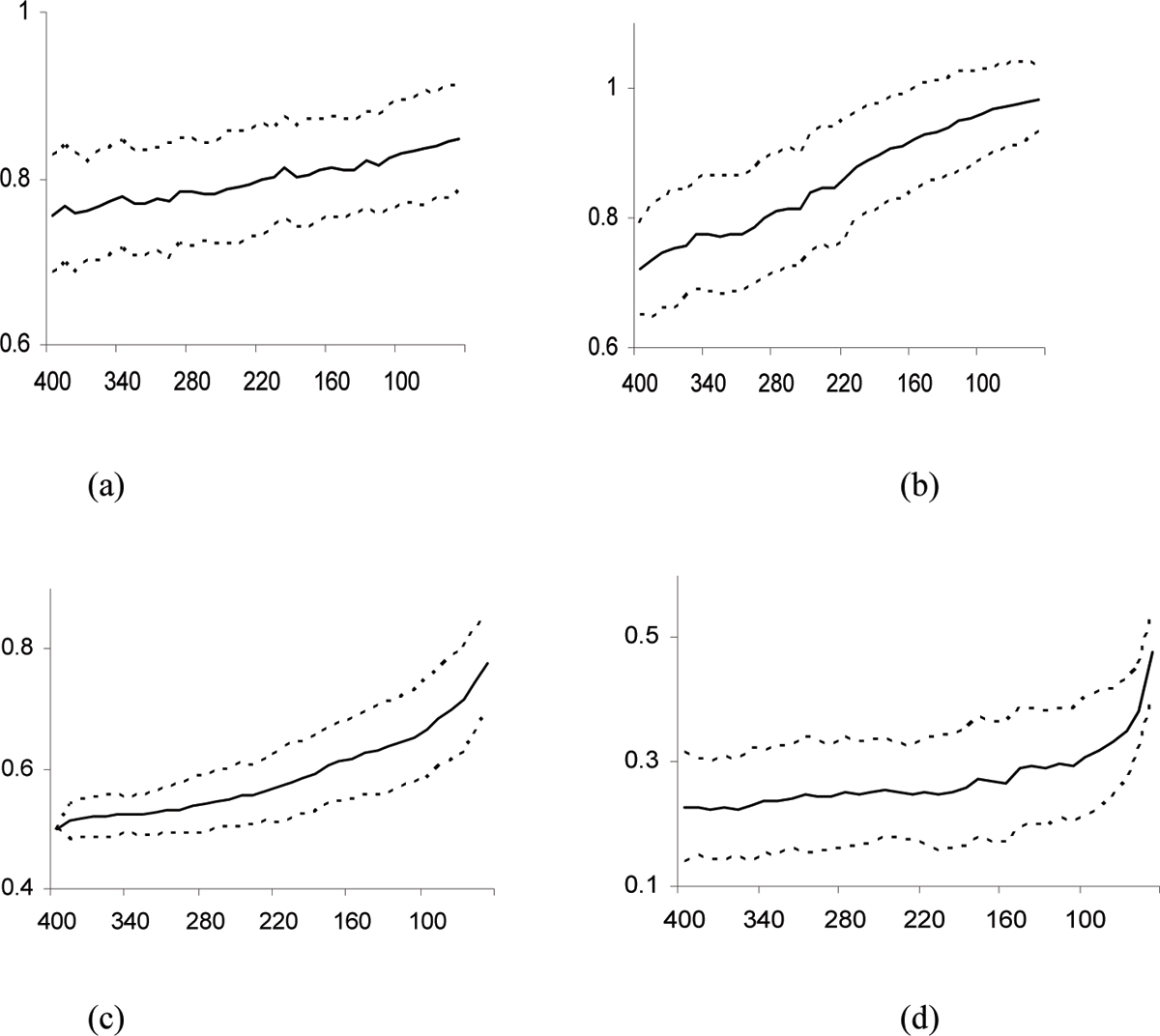}
\caption[The performance analysis on simulated datasets.]{The performance analysis on simulated datasets. The solid lines indicate the average values; and dotted lines indicate
one standard deviation from the averages. The X-axis represents the number of genes in $S_i$. Note that, when the generating process goes, the number of genes in $S_i$ decreases.
(a) The classification accuracy $\theta_{test}$ on the test samples. (b) The classification accuracy $\theta_{train}$ on the training samples. (c) The percentage $\rho_{sim}$ (i)
of truly discriminative genes in $S_i$; (d) The $p$-value $\rho_S(i)$ based on $\hat{\delta}(A_i)$.} \label{fig:diff_cluster_1}
\end{figure}

We run our algorithm with the input gene set contains all the 400 genes for each of the 100 simulated datasets. The performance results are summarized in Figure
\ref{fig:diff_cluster_1}. We observed that the classification performance of the generated clusters keeps increasing as the iteration process goes. The average classification
accuracy of these tests jumps from 0.756 up to 0.848 (Figure \ref{fig:diff_cluster_1}a); and the classification accuracy $\theta_{test}$ on training samples goes up from 0.720 to
0.984 (Figure \ref{fig:diff_cluster_1}b).

We also observed that more and more truly discriminative genes are identified in the active clusters as the algorithm proceeds. Since the genes in the discriminative clusters are
known in each simulated dataset, we computed the ratio $\rho_{sim}(i) = \frac{|S_i \cap (C_1 \cup C_2)|}{|S_i|}$ of the truly discriminative genes over all the genes in for each
iteration $i$. The active clusters output $\rho_{sim}(i)$, just before the algorithm terminates is about 0.778 (Figure \ref{fig:diff_cluster_1}c). Recall that, at each iteration
$i$, the algorithm generates $\kappa=50$ active gene clusters since the number of training samples $n_r = 50$ for each simulated dataset.  We found that at each iteration $i$, the
centroids of two active clusters are very close to $\chi(C_1)$ and $\chi(C_2)$, the centriods of the discriminative clusters in the model. This is reflected by the
indistinguishably small $p$-value $\rho_S(i)$ calculated based on $\bar{d}(A_i, \Delta')$. Here $\bar{d}(A_i, \Delta')$ is the 'average' Euclidean distance of centroids between an
active cluster in $A_i$ and its closest cluster in $\Delta' = \{ C_1, C_2 \}$.

In the same time, the centroids of active clusters become more and more distinguishable from each other, increasingly close to the average pairwise distance of all 400 genes, and
such trend can also be reflected by the increasing $p$-value $\rho_S(I)$ from 0.228 up to 0.476 (Figure \ref{fig:diff_cluster_1}d), calculated based on $\hat{\delta}(A_i)$, the
average Euclidean distance between the centroids of active clusters in $A_i$. Meanwhile, the Silhouette width $\bar{\omega}(A_i)$ of active clusters in $A_i$ increases from 0.826
to 0.980.

\subsection{Leukemia dataset}
Leukemia AML/ALL dataset \cite{golub1999mcc} contains the expression values of 6,817 human genes in 47 acute lymphoblastic leukemia (ALL) and 25 acute myeloid
leukemia (AML) tissue samples. After performing the threshold filtering and logarithmic transformation procedure, we obtained a reduced dataset with only 3,571
genes. Here, we validate our algorithm by using three-fold cross validation. In each run, we randomly selected two third of the samples as the training samples
and the rest as the testing samples. The samples of different phenotype classes are kept proportional in the training and test samples.  The resulting dataset
was further normalized by rescaling the variance of expression values of each gene to 1 in the training samples, and then applying the same rescaling factor to
the expression values of that gene in the test samples.

We conducted the three-fold cross validation for 100 times. To reduce computational cost, we restrict our algorithm on small portions of discriminative genes.
In each run, the algorithm starts with the input gene set consisting of the 357 genes (10\% of all the 3,571 genes) that are highly correlated with the
training samples' phenotype classification in terms of the correlation metric proposed in \cite{golub1999mcc}.

\begin{figure}[tph]
\centering
\includegraphics [width=\columnwidth, viewport=1 1 1200 756, clip] {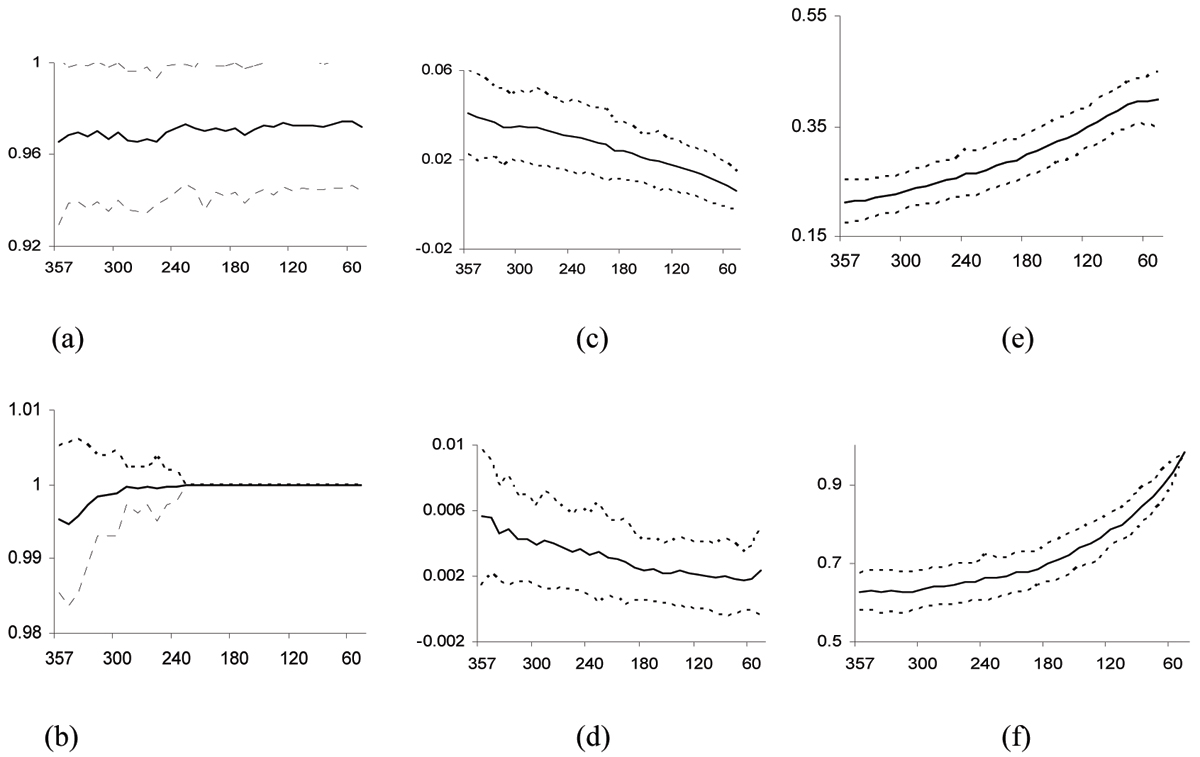}
\caption[The analysis of the three-fold cross validation performance of the algorithm on the Leukemia dataset.]{The analysis of the three-fold cross validation performance of the
algorithm on the Leukemia dataset. The dotted lines indicate the performance values in individual tests. The solid lines indicate the average values; and the dotted lines indicate
one standard deviation from the averages. The X-axis represents the number of genes in $S_i$. (a) The classification accuracy $\theta_{test}$ on the test samples. (b) The
classification accuracy $\theta_{train}$ on the training samples. (c) The $p$-values $\rho_S(i)$ based on $\bar{d}(A_i, \Delta'1)$. (d) The $p$-values $\rho_S(i)$ based on
$\bar{d}(A_i, \Delta'2)$. (e) The $p$-value $\rho_{all}(i)$ based on $\hat{\delta}(A_i)$. (f) The average Silhouette width $\bar{\omega}(A_i)$ of active clusters in $A_i$.}
\label{fig:diff_cluster_2}
\end{figure}

Figure \ref{fig:diff_cluster_2} summarizes the values of the different performance indicators. The average classification accuracy $\theta_{train}$ on the training samples ranges
from 0.994 up to 1 (Figure \ref{fig:diff_cluster_2}b); and the average classification accuracy $\theta_{test}$ on the test samples increase slightly from 0.966 to 0.972 (Figure
\ref{fig:diff_cluster_2}a). These results show that the centroids of the clusters generated in different iteration steps discriminate the training samples better and better
without significant decrease of its generalization ability.

For the evaluation of our algorithm, we searched for perfect 3-gene subsets, which can be used to perfectly classify all 72 samples using the weighted voting classifier trained on
all the samples. This search resulted in 9,722 perfect subsets. We selected 48 (roughly equal to $n_r$) genes $g_i$ ($1 \leq i \leq 48$) with highest occurrence frequency to form
the cluster set $\Delta'_1=\{\{g_i\} | 1\leq i \leq 48\}$ for comparison with the clusters generated by our algorithm.

We also evaluate our algorithm using another cluster set $\Delta'_2$, the final set of active clusters generated by our algorithm with $S'$ as the input gene
subset and with all the 72 samples as the training samples, where $S'$ is the set of the 357 genes (10\% of all the 3,571 genes) that are highly correlated
with the AML/ALL sample classes in terms of the correlation metric proposed in \cite{golub1999mcc}.

Probably because of the selection sensitivity of the correlation metric of \cite{golub1999mcc} resulting from small sample size, the gene sets that are selected according to
different training-test splits do not have many genes in common. In all the 100 validation experiments, only 120 genes appearing in every input gene set  . This number is quite
small compared with 1,071, the number of the genes appearing in some input gene sets (each of size 357).  By contrast, the centroids of clusters in the set   generated in each of
iterations of our algorithm in different runs are significantly similar to the selected discriminative genes in $\Delta'_1$  and $\Delta'_2$ at most iteration steps. This is
reflected by the very small $p$-values $\rho_s(i)$  computed based on $\bar{d}(A_i, \Delta'_1)$  and  $\bar{d}(A_i, \Delta'_2)$, which range from $4.11\times 10^{-2}$ to
$6.12\times 10^{-3}$ (Figure \ref{fig:diff_cluster_2}c) and from $5.62\times 10^{-3}$ to $2.38\times 10^{-3}$ (Figure \ref{fig:diff_cluster_2}d) respectively. The above
observation strongly suggests the stability associated with discriminative clusters rather than with individual discriminative genes. Such stability is one of the main advantages
of our method.

\begin{figure}[tph]
\centering
\includegraphics [width=\columnwidth, viewport=1 1 1200 814, clip] {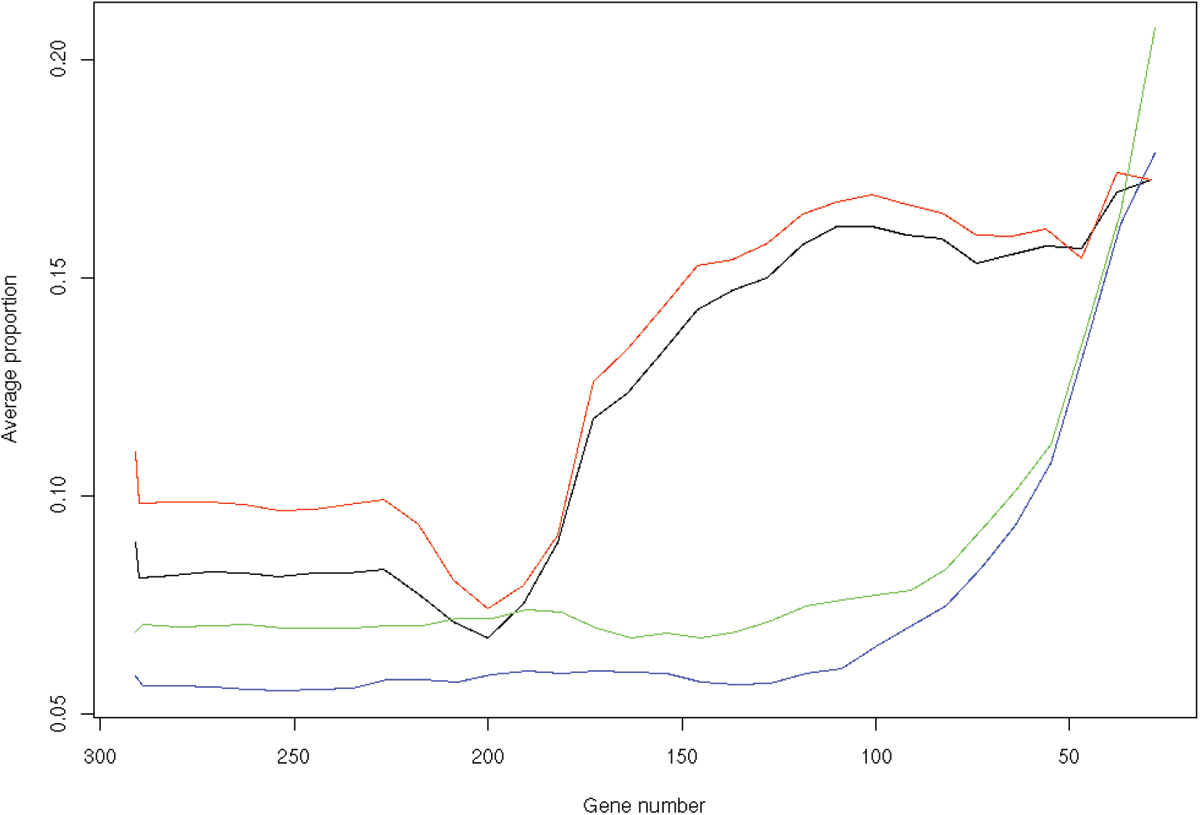}
\caption[Average proportion of genes of the four biological processes during the refinement iterations.]{Average proportion of genes of the four biological processes during the
refinement iterations. The solid black, blue, red, and green lines corresponding to the ordered processes in Table \ref{tab:diff_cluster_1} from top to bottom.} \label{fig:diff_cluster_3}
\end{figure}

We further studied the biological function of genes in the active clusters using Gene Ontology (GO), focusing on the biological processes located at the fifth level of the GO
hierarchy. For the set of all genes from active clusters in $A_i$, we find its enriched biological processes by calculating the hyper-geometric $p$-value, then convert the
$p$-value into a log score $s$ by $s=-\log_{10}(p)$. Table \ref{tab:diff_cluster_1} gives the top four biological processes that are most significantly enriched in the active
clusters in the final iteration, in terms of the score averaged from the 100 validation experiments. All four processes are frequently associated with leukemia. In addition, we
inspected the change of proportion of the genes of the four processes in the active clusters during refinement iterations. The proportions are also averaged over the 100
validation experiments. Figure \ref{fig:diff_cluster_3} shows that when the active clusters contain less than two third genes in input gene set $S$, the average gene proportions
of all four processes monotonically increase until the last iteration. Such convergence strongly suggests that our method can indeed refine clusters into biologically meaningful
ones. Interestingly, processes of inflammatory response and response to wounding showed very similar convergence patterns. In fact, these two processes are closely related. The
same holds for biological processes of regulation of catalytic activity and positive regulation of metabolic process.

\begin{table}
\begin{center}
\begin{tabular}{|l|l|}
  \hline
  Biological process & Average score \\ \hline
  Inflammatory response & 2.72 \\
  Regulation of catalytic activity & 2.19 \\
  Response to wounding & 1.98 \\
  Positive regulation of metabolic process & 1.95 \\
  \hline
\end{tabular}
\end{center}
\caption[Significantly enriched biological processes]{Significantly enriched biological processes}
\label{tab:diff_cluster_1}
\end{table}

\subsection{The performance analysis on other real datasets}
Besides the above dataset, we also tested our algorithm on seven other datasets. The descriptions of these datasets are as follows. Altogether, we derive 12
sample phenotype classification studies from the 8 datasets.

\begin{enumerate}
\item ALL T/B Cell dataset. The 47 ALL samples in Leukemia ALL/AML dataset in
\cite{golub1999mcc} are further divided into 39 T-cell samples and 9 B-cell
samples.

\item Breast cancer dataset \cite{west2001pcs}. This dataset comprises 7,129 genes and 49
samples being divided into two sample phenotype classes according to their estrogen receptor (ER) responses: 25 for ER positive and 24 for ER negative.

\item Carcinoma dataset \cite{notterman2001tge}. It contains the expression levels of about 6600
genes in 18 tumor and 18 normal tissues.

\item Colon dataset \cite{alon1999bpg}. It consists of the expression levels of 6,500 human
genes in 40 tumor and 22 normal tissues.

\item Diffuse Large B-Cell Lymphoma (DLBCL) dataset \cite{shipp2002dlb}. It consists of
expression values of 6,817 genes in 58 DLBCL and 19 Follicular Lymphoma.

\item The Melanoma dataset \cite{bittner2000mcc}. The dataset consists of the expression ratios
of 6,971 human genes in 12 Unclustered Cutaneous Melanomas and 19 Cutaneous
Melanomas samples.

\item Prostate dataset \cite{singh2002gec}. The dataset consists of the expression levels of
52 prostate and 50 normal samples of 6,744 human genes.

\item The Small, round blue cell tumors (SRBCT) dataset \cite{khan2001cad}. It consists of
88 samples divided into five sample phenotype classes: neuroblastoma (NB) (18 samples), rhabdomyosarcoma (RMS) (25 samples), Burkitt lymphomas (BL) (11
samples), the Ewing family of tumors (EWS) (29 samples) and others (5 samples). The dataset has 2,308 genes. Since we consider the binary sample phenotype
classification problem, we derived four binary sample phenotype classification datasets from this dataset using the one-against-all rule: SRBCT-NB, SRBCT-RMS,
SRBCT-BL SRBCT-EWS. RFERENCES
\end{enumerate}

We preprocessed each dataset by applying filtering and logarithm transformation if necessary. For each sample phenotype classification study, we run our
algorithm 100 times by choosing random training-test splits in the same way as the Leukemia dataset described in the last subsection. The performance of our
method is summarized in Table \ref{tab:diff_cluster_2}. In the table, there are two columns for each performance measure, indicating the average values of the
corresponding measures at the first and last iteration step of our algorithm.  Because the exhaustive search of the most frequent globally optimal genes for
constructing $\Delta'_1$ is time-consuming, we only compare the active clusters with $\Delta'_2$  constructed as follows: 1) we apply our algorithm on all
samples in the dataset and 2) use the active clusters of the last iteration as $\Delta'_2$.

\begin{table}
\begin{center}
{\fontsize{9}{9pt}\selectfont
\begin{tabular}{|p{0.1\columnwidth}|l|l|l|l|l|l|l|l|}
  \hline
Datasets    &   \multicolumn{2}{|c|}{$\theta_{test}$}   &    \multicolumn{2}{|c|}{$\rho_S(i)$ based on $\bar{d}(A_i, \Delta'_2)$}   &    \multicolumn{2}{|c|}{$\rho_{all}(i)$ based on $\bar{\delta}(A_i)$}   &    \multicolumn{2}{|c|}{$\bar{\omega}(A_i)$}  \\ \hline
Leukemia ALL T/B cell   &   0.97    &   0.977   &   1.72E-02    &   8.28E-03    &   0.088   &   0.331   &   0.406   &   0.973   \\ \hline
Breast  &   0.843   &   0.842   &   1.33E-02    &   8.36E-03    &   0.142   &   0.421   &   0.351   &   0.974   \\ \hline
Carcinoma   &   0.983   &   0.981   &   2.96E-02    &   3.20E-02    &   0.194   &   0.252   &   0.382   &   0.966   \\ \hline
Colon   &   0.814   &   0.806   &   2.43E-02    &   2.06E-02    &   0.75    &   0.755   &   0.673   &   0.978   \\ \hline
DLBCL   &   0.896   &   0.929   &   8.75E-02    &   1.99E-02    &   0.441   &   0.514   &   0.716   &   0.982   \\ \hline
Melanoma    &   0.913   &   0.921   &   1.71E-02    &   2.25E-02    &   0.129   &   0.463   &   0.272   &   0.957   \\ \hline
Prostate    &   0.889   &   0.916   &   4.79E-02    &   2.27E-02    &   0.495   &   0.541   &   0.68    &   0.987   \\ \hline
SRBCT-BL    &   1   &   1   &   3.63E-04    &   7.52E-05    &   0.314   &   0.322   &   0.682   &   0.984   \\ \hline
SRBCT-EWS   &   0.956   &   0.986   &   5.06E-04    &   9.17E-05    &   0.297   &   0.408   &   0.634   &   0.984   \\ \hline
SRBCT-NB    &   0.989   &   0.996   &   2.99E-04    &   6.42E-05    &   0.321   &   0.436   &   0.665   &   0.986   \\ \hline
SRBCT-RMS   &   0.974   &   0.98    &   4.82E-04    &   8.18E-05    &   0.304   &   0.347   &   0.63    &   0.989   \\ \hline
Lukemia AML / ALL   &   0.966   &   0.972   &   5.62E-03    &   2.38E-03    &   0.212   &   0.398   &   0.627   &   0.98    \\ \hline
\end{tabular}
}
\end{center}
\caption[The performance of the algorithm for different sample phenotype classification studies.]{The performance of the algorithm for different sample
phenotype classification studies.} \label{tab:diff_cluster_2}
\end{table}

The classification accuracy $\theta_{test}$ on the test samples shows that among 9 of 12 sample phenotype classification studies, the prediction performance of
active clusters in $A_j$ increases slightly from the start to the end of each execution, which are highlighted in the table. The value of $\theta_{test}$ for
the remaining three studies (Breast, Colon and Carcinoma) decrease slightly. The above observations indicate that for all datasets we tested, there is no
significant decrease in the generalization ability of the active clusters $A_i$ in obtained in each iteration step. The classification performance
$\theta_{train}$ on the training samples increases in all of the 12 studies, which indicates that the separation of the training samples improves for all
studies.

All the 100 input gene sets $S$ vary a lot in different runs for each study. There are only 1.1\% to 5.1\% of all the genes appearing in all the 100 input gene sets $S$, while at
least 23.8\% to 51.7\% genes appear in some input gene sets. By contrast,  the centroids of clusters in $SA_i$ generated by our algorithm at each iteration step $i$ are stably
close to the optimal centroids of clusters in $\Delta'_2$ as reflected by the $p$-values $\rho_s(i)$ ranging from $2.99\times 10^{-4}$ to $8.75\times 10^{-2}$ at the first
iteration step and those ranging from $6.42\times 10^{-5}$ to $3.20\times 10^{-2}$ in the last iteration step. The consistent closeness of the clusters generated in different
repeats can also be reflected in the standard deviation of $\rho_s(i)$, which are limited from 0.32 to 0.96 times of the absolute values of   in the first iteration step and 0.24
to 1.37 times at the last iteration step.

During the generation process, the $p$-values $\rho_{all}(i)$ of average pairwise distance $\hat{\delta}(A_i)$ among centroids of clusters in $A_i$ keeps increasing for all 12
studies (ranging from 0.088 to 0.750 at the first iteration step and from 0.252 to 0.755 in the last step), and the average Silhouette width of active clusters $\bar{\omega}(A_i)$
keeps increasing for all the 12 studies (ranging from 0.230 to 0.698 at the first iteration step and from 0.964 to 0.989 in the last iteration step). This indicates that the
clusters in $A_i$ are more and more distinct in general.

In summary, our test shows that on real microarray datasets, our algorithm is able to generate clusters that separate the training samples with increasing prediction accuracy and
closeness to known optimal clusters. Such discriminative cluster refinement is consistent with what we have observed on simulated datasets.

\subsection[Comparing the sample phenotype classification performance to other studies]{Comparing the sample phenotype classification \\ performance to other studies}
In this section, we compare the cross validation performance of our method with previous works
reported in \cite{jornsten2003sgc, dettling2002scg, shipp2002dlb, jager2002igs}. For the purpose of
comparison, we converted the classification performance from the classification accuracy
$\theta_{test}$ into the error rate. Table \ref{tab:diff_cluster_3} summarizes the comparison of
our algorithm (of both binary and multi-sample phenotype class versions) with others by the cross
validation error rates. It is difficult to make direct comparisons with other approaches in the
literature, because the specific data sets or data preparation are not always available. However,
the performances our method is in general comparable to others. In the comparison, the DLBCL and
Carcinoma datasets are validated using leave-one-out validation; and the remaining datasets are
validated using three-fold cross validation.

\begin{table}
\begin{center}
{\fontsize{9}{9pt}\selectfont
\begin{tabular}{|p{0.2\columnwidth}|p{0.15\columnwidth}|p{0.15\columnwidth}|p{0.15\columnwidth}|p{0.15\columnwidth}|}
  \hline
  Datasets & Our algorithm & Dettling and Buhlmann (2002) & Jornsten and Yu (2003) & Shipp et al. (2002) \\ \hline
  Lukemia AML/ALL & 3.43 - 2.57 & 6.58 - 2.71 & & \\ \hline
  Leukemia three classes & 13.8 - 9.3 & & 12.6 & \\ \hline
  Breast & 16.14 - 14.11 & 3.00 - 0.75 & & \\ \hline
  Carcinoma & 5.6 - 0.0 & & & \\ \hline
  Colon & 19.41 - 18.23 & 23.35 - 15.95 & 13.6 & \\ \hline
  DLBCL & 8.7 - 7.4 & & & 7.8 \\ \hline
  Prostate & 11.09 - 8.36 & 16.47 - 6.91 & & \\ \hline
  SRBCT multi class & 5.92 - 4.27 & 5.76 - 0.43 & & \\ \hline
\end{tabular}
}
\end{center}
\caption[Comparison of our algorithm with others.]{Comparison of our algorithm with others.}
\label{tab:diff_cluster_3}
\end{table}

Dettling and Buhlmann \cite{dettling2002scg} reported the error rate of their algorithm for different datasets. They employed nearest neighbors and aggregated trees as the
classifiers in their three-fold cross validation test. For the leukemia AML / ALL dataset, our algorithm seems to achieve a slightly lower error rate than theirs. In the Colon and
Prostate datasets, the error rate of our algorithm lies between that of theirs. For the Breast dataset, the error rate is significantly higher than that of Dettling and
Buhlmann's. However, we obtained the performance using all the original 49 samples. The error rate in each test ranges from 7.89 to 6.90. According to \cite{west2001pcs}, at least
7 out of the 49 samples are inherently erroneous. In comparison, Dettling and Buhlmann \cite{dettling2002scg} used the 38 good samples selected by \cite{west2001pcs}, and the
error rate ranges from 1.14 to 0.10. The 38 samples used in Dettling and Buhlmann \cite{dettling2002scg} consists none of the above 7 erroneous samples. Thus, we believe that the
performances of ours and Dettling and Buhlmann's are still comparable for the Breast dataset.

For the DLBCL dataset, the leave-one-out performance of Shipp et al. \cite{shipp2002dlb} is in our performance range. For Carcinoma dataset, Jaeger et a. \cite{jager2002igs}
achieved perfect leave-one-out performance, and our best performance can match theirs.  For the Colon dataset, both ours and Dettling and Buhlmann's error rate are higher than
Jornsten and Yu's.

We also test the performance of the multiple-sample phenotype class version of our method against other methods. For the Leukemia three-class dataset, our
method is comparable to Jornsten and Yu's method. However, for the SRBCT multi class dataset, our algorithm seems had a slightly higher error rate than that of
Dettling and Buhlmann's.

\section{Conclusions}
Due to the small-sample-high-dimension nature of the microarray dataset, it is not difficult to find highly discriminative gene subsets of small size. However,
if a gene selection process is unstable with the choice of training samples, the biological significance of the resulting gene subsets is often not guaranteed.
In this paper, instead of finding individual discriminative genes or gene subsets, we propose a novel backward approach for generating a series of highly
discriminative gene clusters. Compared to selection of individual discriminative genes, genes grouped in these clusters are more stable when subject to change
of training samples. Therefore they could provide more convincing support to gene interactions that are associated with the sample phenotype classes. In
future, we will work with biologists to study the biomedical implication of these clusters.

Regarding to the sample phenotype classification performance, the gene clusters produced by our approach can generally achieve good cross validation
performance compared to the existing methods for most of datasets we tested. More importantly, our test experiments show that regardless of the choice of
training samples, the centroids of the clusters generated are stable and significantly close to the known optimal gene clusters found using all the samples.
All these indicate that our approach is promising. However, the current version of our algorithm is time-consuming. In future, the computational efficiency
will be investigated. On the other hand, we used K-means algorithm, a typical partitioning based clustering method to seek a certain number of clusters that
minimize the sum of squared distances between each gene and its centroid. The drawback for K-means is the subjective specification of input parameters such as
the number of clusters and initial centroid locations. For unknown microarray datasets, such information is unavailable. Furthermore, different input
parameters may result in significantly different clustering results. K-means can only converge to local optima, rather than the global optimum. In order to
address these problems associated with K-means clustering. We plan to apply a novel clustering method based on Random Matrix Theory (RMT) \cite{zhu2008pca}
which is completely objective and do not require the specification of cluster number and initial centroid locations. RMT method avoids being trapped into local
optima. Furthermore, most previous clustering methods including K-means and hierarchical clustering partition members into non-overlapping groups. The RMT
method allows the same genes in multiple groups to reflect the fact that a single gene may contribute to multiple biological pathways.

In order to test the discriminative power of a certain gene cluster, additional criteria established by statistical analysis should also be conducted to identify and remove
inactive cluster. For example, gene expression pattern observed in the active clusters should be less likely to appear in the control set. Chi Square test might be used to test
the significance. Some data normalization technique may be considered in the preprocessing step to improve the data quality. Furthermore, more suitable backward feature selection
method needs to be exploited so that the gene clustering and cluster selection processes can be integrated better. Our approach provides a flexible framework that allows us to
test the performance of various computing modules in a various ways of combinations.

Our method can not only be easily extended to multi-sample phenotype class prediction, but also can be easily extended to integrate gene clusters obtained from
multiple datasets that contain sample phenotype classification study of same type. Such integration can be obtained by counting the occurrence frequency of a
gene in similar clusters obtained from different datasets. In such way, by pulling evidences from multiple datasets from different experimental sources, genes
in such clusters of high occurrence frequency would have higher confidence to form truly discriminative gene sets.

\section{Methods}
\subsection{Algorithm}
In this subsection, we present our backward approach for generating discriminative gene clusters. The method is executed in a repetitive manner. In each pass,
the method first groups genes into clusters that may indicate functional categories \cite{tavazoie1999sdg}. It then ranks the generated clusters and eliminates
those clusters that are less discriminative so that the re-clustering of remaining genes can generate modules with better resolution and stronger association
with the sample phenotype classes. In the clustering stage, we use the K-means method to group the genes into a constant number of active clusters.

In the elimination stage, we use a backward feature selection method. This stage involves cluster validation and evaluation of the discriminative ability of active clusters. To
validate clusters, we use the Silhouette width \cite{kaufman1990fgd} to measure their validity. Assume the input genes are partitioned into $p$ clusters $C_i, C_2, \ldots, C_p$.
Given a gene $g$, let $\bar{w}_g$ be the average Euclidian distance between $g$ and another gene within the same cluster, and let $\bar{b}_{gJ}$ the average Euclidian distance
between and a gene $g$ in a different cluster $C_J$. Then the Silhouette width $\omega_g$ of $g$ is defined as $\omega_g = \frac{\min_J(\bar{b}_{gJ}) -
\bar{w}_g}{\max(\min_J(\bar{b}_{gJ}), \bar{w}_g)}$, and the Silhouette width of a cluster is defined as the average Silhouette width of all its members. It is easy to see that the
Silhouette width of a cluster fall within the range from -1 to 1. A good cluster should have a high Silhouette width.

To measure the discriminative ability of an active cluster, we adopt the idea of SVM-RFE method in \cite{guyon2002gsc}. Support Vector Machine (SVM) is a binary-class prediction
method originated from statistical learning theory \cite{vapnik2000nsl}. A linear SVM first finds a decision hyperplane $y = \mathbf{w}^T\mathbf{x} + b$ that maximizes the
separation between samples of two classes; and then it does class prediction according to the relative location of a new sample with respect to the hyperplane in the feature
space. Note that the weight vector $\mathbf{w}$ found by the linear SVM indicates the relative importance of the genes for the classification.  Here, we iteratively train a linear
SVM and eliminate a gene cluster based on an overall evaluation on both the weight and the Silhouette width instead of discarding single gene in the original SVM-RFE method. Such
systematic approach makes the elimination process to better reflect the underlying biological meaning.

\begin{figure}[tph]
\centering
\includegraphics [width=\columnwidth, viewport=1 1 1200 749, clip] {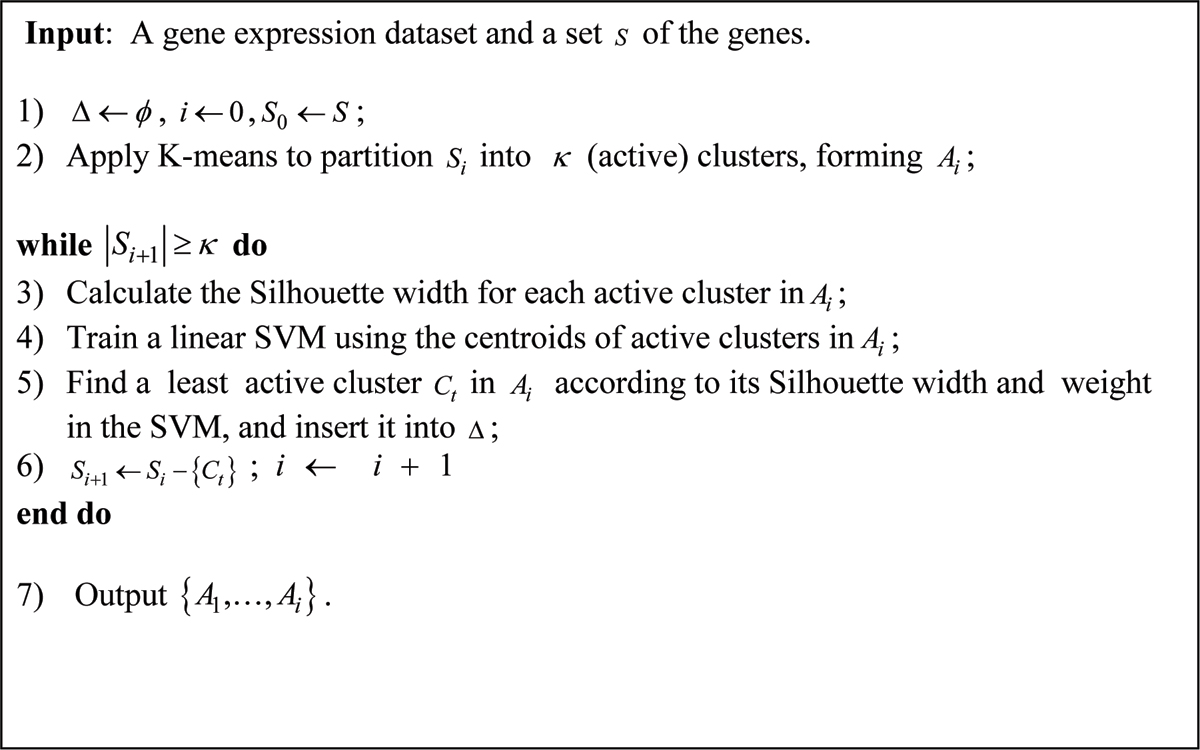}
\caption[The algorithm of selecting discriminative clusters.]{The algorithm of selecting discriminative clusters.} \label{fig:diff_cluster_4}
\end{figure}

Our method is summarized into the algorithm in Figure \ref{fig:diff_cluster_4}. In the algorithm, $\Delta$ denotes the set of inactive gene clusters;  $A_i$ denotes the set of
active clusters at each iteration i; $S_i$ denotes the set of genes under consideration at the beginning of the iteration $i$; $\kappa$ denotes the number of clusters partitioned
at each iteration step. For simplicity, we set $\kappa$ to be $n_r$, the number of training samples.

It is often difficult to determine how many clusters the genes should be grouped into for microarray datasets, which usually have complex expression patterns. The algorithm
outputs $\kappa=n_r$, the number of the training samples, active clusters in each iteration. This is because the expected size of a feature subset that can linearly discriminate
all the samples is only $(n_r + 1) / 2$ \cite{xu2003gsf}. Note that if the feature number is too small, the clustering will lose its resolution.

Recall that the K-means clustering method starts with an initial partition of the genes. In order to make it more deterministic in Step 2, we first select $\kappa$ genes as
follows: Find a furthest gene pair and form an initial gene set $G$, and then iteratively find a gene with largest average Euclidean distance from the genes in $G$ and add it into
$G$ until $|G|=\kappa$. We then partition all the genes into clusters by merging each gene with its nearest gene in $G$.

The calculation of the Silhouette width of each cluster in $A_i$ takes all the clusters in both sets $A_i$ and into account $\Delta$. At the $i$th iteration, the algorithm groups
all the genes in the set $S_i$ into $\kappa$ clusters, forming the cluster set $A_i$, and then insert the least active cluster into the inactive cluster set $\Delta$ in Step 5 as
follows.

There are two important factors to evaluate in order to determine which cluster should be removed from $S_i$ and added into $\Delta$. The first factor is the cluster's Silhouette
width. Another factor is the cluster's discriminative ability in terms of its weight determined by the linear SVM constructed in Step 4. Here, we would like to eliminate a least
discriminative cluster whose centroid is sufficiently representative of the expression pattern of the cluster (measured by the Silhouette width). In other words, we eliminate a
set of well clustered genes whose expression patterns have little contribution to classification. On the other hand, those not well clustered genes will be re-clustered at later
iterations.

\begin{figure}[tph]
\centering
\includegraphics [width=\columnwidth, viewport=1 1 1200 340, clip] {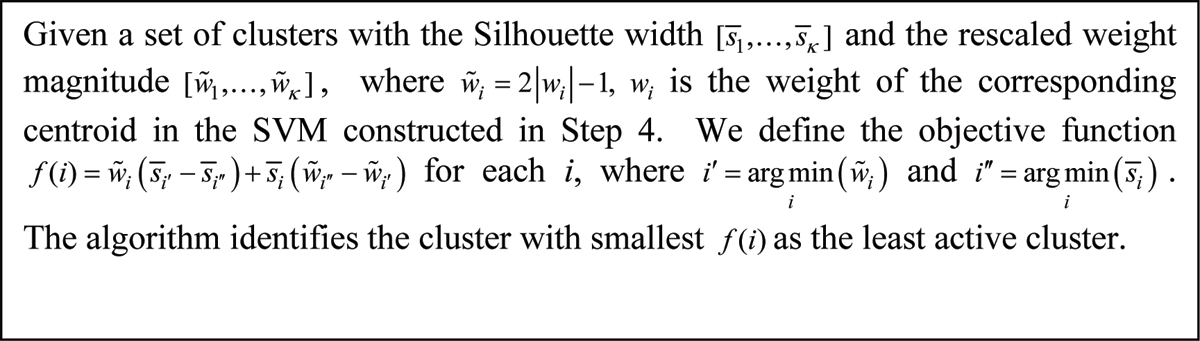}
\caption[Multiple objective optimization procedure for cluster elimination.]{Multiple objective optimization procedure for cluster elimination.} \label{fig:diff_cluster_5}
\end{figure}

Since the above two factors are not always consistent, we adopt a multiple objective optimization technique appearing in \cite{gen14ecm} to find a nice tradeoff between these two
factor and such multiple objective method is shown in Figure \ref{fig:diff_cluster_5}.

Finally, we can extend the algorithm to the multiple-class case by adopting the popular one-against-all approach. In this approach, given a training test split, both training and
test samples of a dataset of $K > 2$  classes are transferred into $K$ binary classification problems, each corresponding to classify samples from one class against samples from
all remaining classes. Then our algorithm executed on the $K$ problems results in $K$ series of active cluster sets $A_{j,i},j=1,\ldots,K$. Then classifiers are constructed using
$K\times \kappa$ clusters from the $K$ active cluster sets $A_{1,i_1}, \ldots, A_{K, i_K}$ by selecting $i_1, \ldots, i_K$ such that $|S_{i_1}|, \ldots, |S_{i_K}|$ are roughly
identical. Given the centroids of the above $K\times \kappa$ clusters, a multi-class linear SVM is trained using training samples and tested on test samples.

\subsection{Performance measures}
We validate our method in terms of its classification performance and clustering performance.  The classification performance is determined by the classification accuracy on
training or testing samples. We use the SVM as the classifier to evaluate the generated gene clusters. Classification accuracy $\theta_{test}$ on the test samples is defined as
the percentage of the correctly classified samples. However, we define classification accuracy $\theta_{train}$ on training samples as the average accuracy of the 10-fold cross
validation on the training samples as suggested in \cite{ambroise2002sbg} for less biased estimation of classification performance.

The clustering quality is measured in terms of the density of the clusters, as well as the distinction between clusters and the closeness of the clusters to some reference
clusters. They are measured respectively by (a) the average Silhouette width $\bar{\omega}(A_i)$ of active clusters in $A_i$ produced in iteration $i$, (b) the average Euclidean
distance $\bar{\delta}(A_i)$ between the centroids of active clusters in $A_i$ and (c) the 'average' Euclidean distance $\bar{d}(A_i, \Delta')$ of centroids between an active
cluster in $A_i$ and its closest cluster in a reference cluster set $\Delta'$ (the construction can be found in the Result Section). To be more precise, assume $A_i = \{ C_1,
\ldots, C_\kappa\}$ and $\Delta' = \{ D_1, \ldots, D_\kappa \}$. First, from 1 to $\kappa$, find recursively $C'_l \in A_i$ and $D'_l \in \Delta'$ such that $d(x(C'_l), x(D'_l)) =
\min_{C'_l\in A_i - A'', D''_l \in \Delta' - \Delta''_l}d(x(C''_l), x(D''_l))$, where $x()$ denotes the centeroid of a cluster, $d(,)$ the Euclidean distance between two vectors,
$A'' = \{C'_1, \ldots, C'_{l-1}\}$ and $\Delta''_l=\{D'_1, \ldots, D'_{l-1}\}$. Then, the 'average' Euclidean distance $\bar{d}(A_i, \Delta')$ is defined as $\bar{d}(A_i, \Delta')
= \frac{1}{\kappa} \sum_{i\leq l \leq \kappa} d(x(C'_l), x(D'_l))$.

We measure the statistical significance of average distances in both case (b) and (c) at each iteration $i$ against the pairwise distances of all genes in the input gene set $S$
in terms of the $p$-value $\rho_S(i)$, and against all the genes in the dataset in terms of the $p$-value $\rho_all(i)$. In each case, the $p$-values are calculated according to
the empirical distribution (null distribution) of the pairwise distance of genes randomly sampled in the whole dataset.

\chapter{An integrative approach to characterize disease-specific pathways and their coordination} \label{Chp:NetModule}

\section{Introduction}
The recent development of microarray technology has significantly facilitated the identification of disease-related genes ~\cite{sevenet2003dmc, chen2002gep,
golub1999mcc, liu2006ige}. However, many complex disease phenotypes are caused not by individual genes, but by the coordinated effect of many genes. Insight
into the structure and coordination of disease-related pathways is crucial to understanding the pathophysiology of complex diseases. However, it has proved
difficult to infer pathways from microarray data by deriving modules of multiple related genes, rather than individual genes. The major challenges are: (1)
Genes involved in a pathway may exhibit complex expression relationships beyond co-expression, which may be overlooked by standard microarray analysis methods
such as clustering ~\cite{zhou2005faa}. (2) Pathways are dynamic and the current static annotation of pathways may not serve as a good template. In fact,
pathways are manual dissections of the underlying dynamic gene regulatory network. Under different conditions, different segments of the ensemble network will
be activated, leading to condition-specific activation of pathways ~\cite{yan2007gba}.

\begin{figure}[tph]
\centering
\includegraphics [width=0.7\columnwidth] {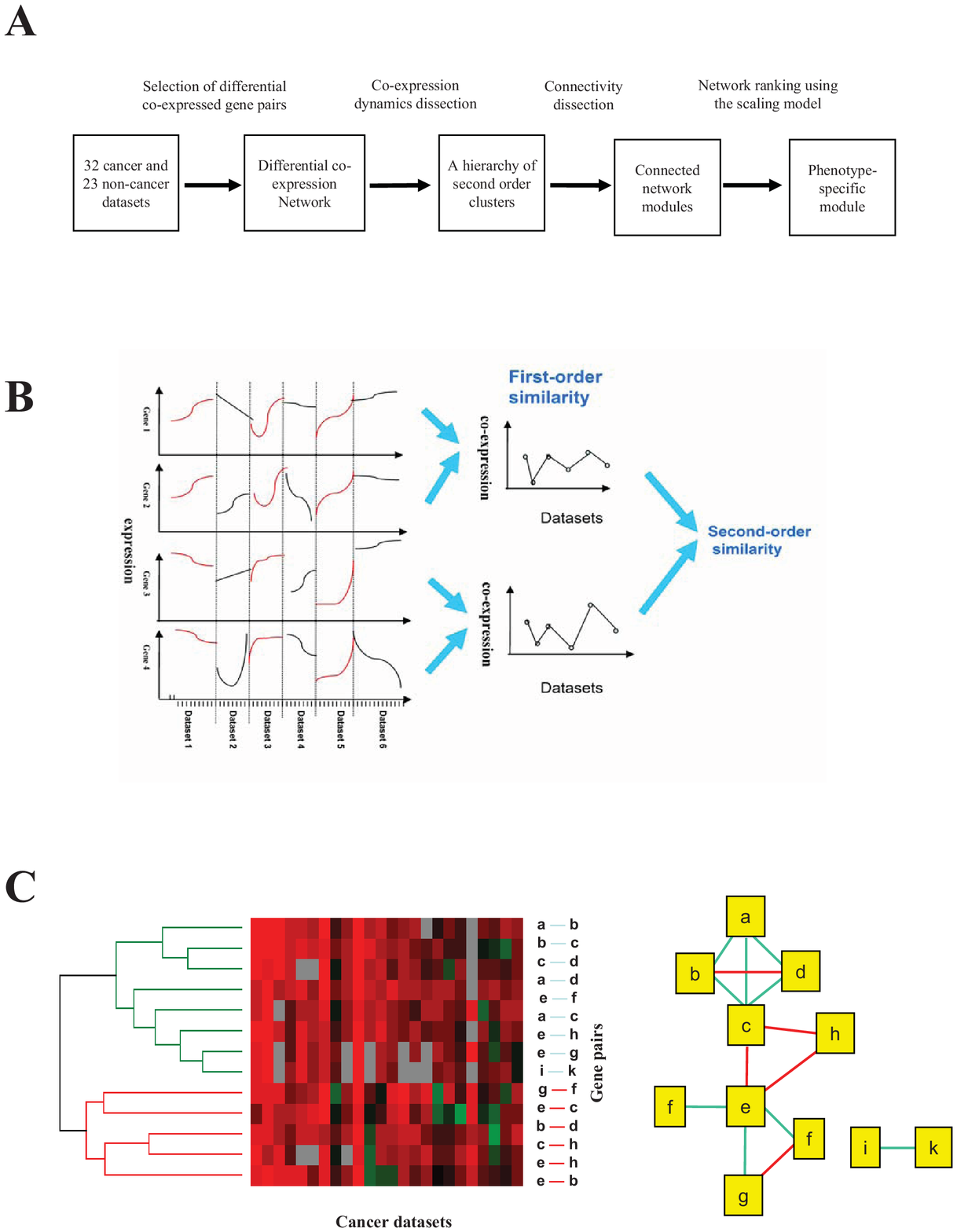}
\caption[Overview of network module finding procedure.]{Overview of analysis procedure. (A) Flow chart of the analysis pipeline. (B) Schematic illustration of
the concept of second-order similarity. It is obvious that the overall expression similarity between the two gene pairs (genes 1 and 2 versus genes 3 and 4) is
not significantly high, but their first-order expression correlation profiles exhibit high second-order similarity. (C) Schematic illustration of the
dissection of differential co-expression networks into network modules based on the co-expression dynamics and network connectivity. In the heat map, every
column corresponds to a dataset and every row corresponds to a gene pair. Red, black, green and grey corresponds to positive, low, negative and missing
correlations, respectively. By hierarchical clustering, the gene pairs fall into two major second-order clusters. The 9 gene pairs in the green cluster
comprise three connected network components, whereas the 6 gene pairs in the red cluster give rise to three connected components. Furthermore, by considering
the second order cluster of a higher level of the hierarchy, which consists of both green and red clusters, the six networks are united to form two connected
networks, reflecting the hierarchical modularity of cancer co-expression networks. } \label{fig:cancer_motif_1}
\end{figure}

In this study, by integrating many microarray datasets we propose a novel method to simultaneously infer pathways and disease/phenotypic conditions under which the pathways are
activated. The identified pathways may comprise genes with complex expression relationships beyond co-expression. Due to the existence of a large amount of cancer microarray data,
we used cancer as our case study. We collected a series of microarray datasets measuring different types of cancers, and a series of datasets measuring other
cellular/physiological conditions. We first construct a differential co-expression network, in which each node represents a gene and each edge indicates a gene pair that is
frequently co-expressed in cancer datasets but not in non-cancer datasets. We then dissect the networks into cancer-subtype specific network modules by considering (1)
co-expression dynamics and (2) network topology.  Figure ~\ref{fig:cancer_motif_1}a illustrates the conceptual pipeline of our method.

To measure co-expression dynamics, we use second-order expression similarity, which we proposed previously ~\cite{zhou2005faa}. Briefly, if we define
first-order expression similarity as the expression similarity of two genes from one dataset, then second-order similarity measures whether two gene pairs
simultaneously exhibit either high or low expression similarity across multiple datasets. In general, high first-order similarity suggests the existence of a
functional link between two genes, and clustering based on the second-order similarity captures multiple functional links always activated and deactivated
under similar conditions. Such functional links are likely to comprise a functional module. Interestingly, genes in a second-order cluster may not always have
high first-order similarity (see an example in Figure ~\ref{fig:cancer_motif_1}b); therefore, second-order analysis allows us to identify functional modules
that are inaccessible to co-expression analysis.

Given multiple gene pairs sharing high second-order similarity, we further divide them into network components based on their connectivity on the differential
co-expression graph (see an example in Figure ~\ref{fig:cancer_motif_1}c). We observe that genes within a connected network component are more likely to
participate in the same specific pathway than those between different components, which, in turn, are likely to be involved in different relevant pathways.
This may reveal high-order cross-pathway coordination. In fact, hierarchical clustering of differentially co-expressed gene pairs based on their second-order
similarity results in a hierarchical modularity in terms of relevance of functional links. We designed a linear scaling model to select modules by considering
both module size (number of edges) and within-module second-order similarity. Then, given selected modules, we can further infer datasets (phenotypic
conditions) in which a module is activated, i.e. in which genes in the module coordinate.

Applying our methods to 32 cancer-related microarray datasets, and 23 non-cancer related datasets, we derived 162 second-order clusters consisting of 224 network modules,
activated either in cancer or in specific cancer subtypes. In particular, we identified a breast cancer specific network module that involved in tumor suppression via
platelet-derived growth factor (PDGF)-like signaling, more importantly, a hub gene PDGFRL that may play an important role in this tumor suppressor module.

\section{Results}
\subsection[Network properties of the cancer differential co-expression network]{Network properties of the cancer differential \\ co-expression network}
We curated 32 human microarray datasets (1,764 expression profiles in total) measuring cancers of
12 tissues, and 23 datasets (1,158 expression profiles) not
related to cancer (e.g. normal tissues, chronic granulomatous disease, Huntington's disease, inflammatory response). 
We first identify gene pairs which consistently demonstrate higher correlation in cancer versus non-cancer datasets based on a
robust correlation estimator, the normalized Percentage Bend correlation (for details see Methods). In following sections, if not specified, the term
correlation will by default refer to the normalized Percentage Bend correlation. These criteria result in 6,035 gene pairs covering 1,967 genes. The 6,035 gene
pairs, each representing a potential conditional functional link, can be represented as a differential co-expression network. In this network, each gene is
represented as a node and each differential co-expression relationship is represented as an edge.

\begin{figure}[tph]
\centering
\includegraphics [width=\columnwidth, viewport=1 1 1373 963, clip] {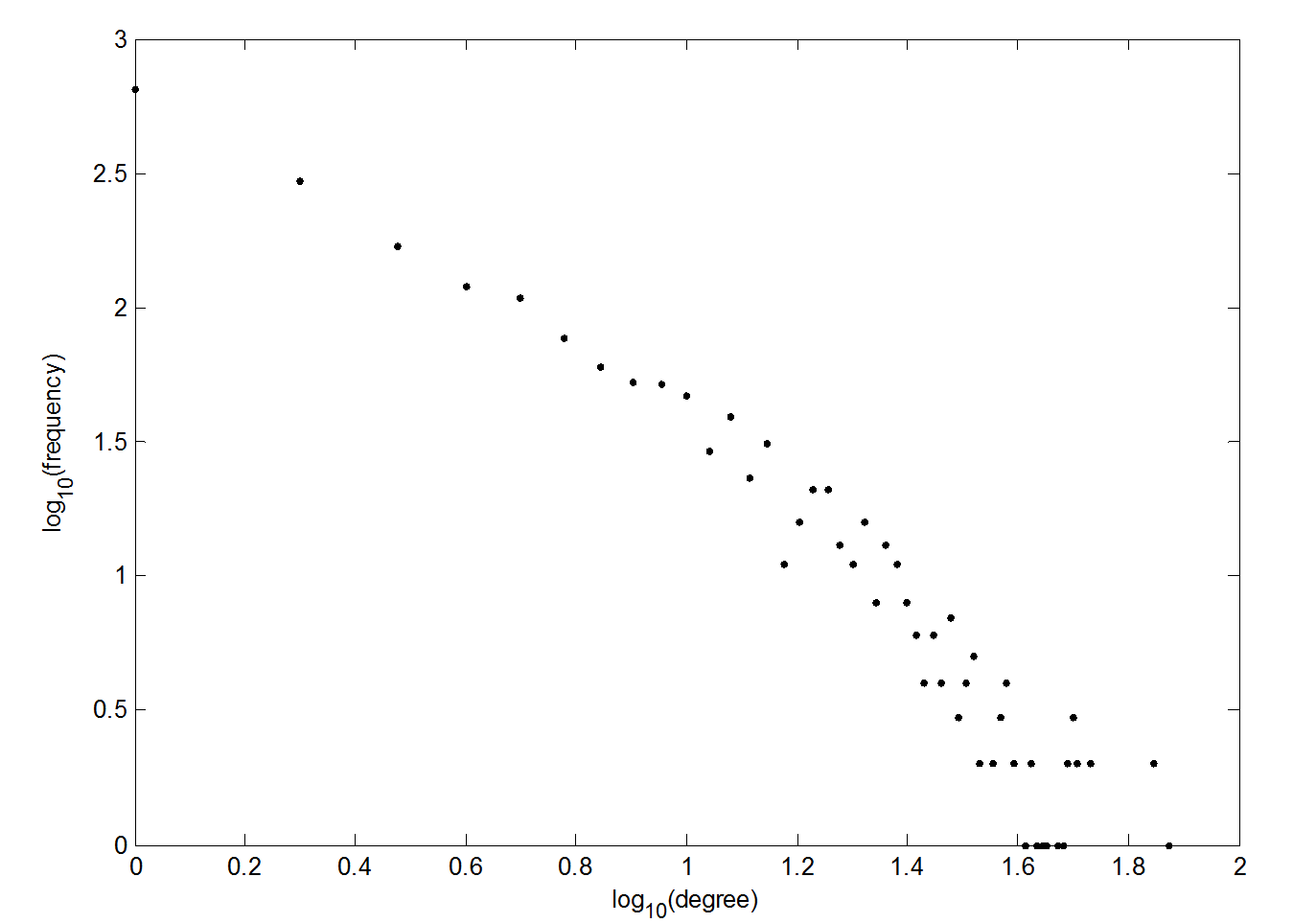}
\caption[Node degree distribution on the differential co-expression
networks.]{Node degree distribution on the differential co-expression
networks.} \label{fig:cancer_hub_gene_degree}
\end{figure}

It has been reported that co-expression networks follow a scale-free node degree distribution ~\cite{jordan2004cac}. We observed that the differential
co-expression network also follows such a topology, where only a small number of nodes act as ``highly connected hubs'' (see node degree distribution in Figure
\ref{fig:cancer_hub_gene_degree}). This indicates that most gene-gene co-expression relationships differing between cancer and other phenotypes are associated
with only a few ``hub'' genes. Such hub genes exhibit a high degree of coordination with many other genes in neoplastic states, and are therefore likely to
play important roles in carcinogenesis and cancer progression. In fact, most hub genes fall into two main functional categories: 1) core processes of
neoplastic states such as cell division and chromosome organization; or 2) dynamic interactions between cancer cells and their microenvironment such as
angiogenesis, immune response, and cell adhesion . For those hub genes with unknown functions, we can predict their cancer-related functions based on their
neighbor genes. For example, the 16 out of the 33 interacting partners of the ADP-ribosylation factor-like 6 interacting protein (ARL6IP) are involved in cell
division (hypergeometric test $p$-value $1.6 \times 10^{-24}$). Thus, ARL6IP is likely to be involved in cell proliferation, consistent with its initial
characterization as an interaction partner of the Ras superfamily member ARL6 ~\cite{pettersson2000ccl}. As another example, while microfibrillar-associated
protein 2 (MFAP2) has long been known to bind to various components of the elastic extracellular matrix ~\cite{jacob2001res}, it has not been clear whether it
serves more than a mechanical function. We found that 6 out of its 24 neighbor genes are involved in cell adhesion ($p$-value $7.7 \times 10^{-5}$). In fact, a
recent study found that MFAP2 binds to a neighbor gene Notch1 and activates it ~\cite{miyamoto2006mpm}.

In the differential networks, we selected the top 112 hub genes that have degrees $\geq$ 22, such that they together account for 30\% of the total degrees on
the differential co-expression networks. 
Many of those genes fall into two main functional categories: 1) core processes of
neoplastic states such as cell division and chromosome organization (36 genes); or 2) dynamic interactions between cancer cells and their microenvironment such
as angiogenesis, immune response, and cell adhesion (24 genes). Genes falling into category (1) include TUBB (participates in microtubule-based movement; node
degree 70), CKS1B (cell cycle; node degree 43), KPNA2 (regulation of DNA recombination; node degree 44), RRM1 (DNA replication; node degree 41), MAD2L1 (cell
cycle; node degree 49), and SOX2 (establishment/maintenance of chromatin architecture; node degree 30). Genes in category (2) include MAP2K7 (response to
proinflammatory cytokines; node degree 50), TYROBP (cellular defense response; node degree 37), CD4 (immune responses; node degree 45). They are all reported
to play a role in cancer pathogenesis and progression. In addition, hub genes in the category (1) behave as hub genes across most of datasets, while those in
the category (2) tend to be in hub genes only in solid tumor datasets.

\subsection{Identification of pathway modules specific to cancer or cancer subtypes}
The differential co-expression network provides a summary of co-expression links frequently active across all types of cancers. However, it does not provide
clues as to which set of links tend to be simultaneously active and inactive under which types of cancer. That is, the edges of a differential co-expression
network may not be active in the same subset of datasets. In fact, the largest connected component of the differential co-expression network contains 5944
edges, which comprises 98\% of all the edges in the network. Thus, based on connectivity alone we cannot break the network into functionally coherent and
cancer-subtype specific modules.

To dissect the networks, we integrate two types of information, the co-expression dynamics and the network connectivity, to extract cancer-subtype specific
network modules. First, we employ the second-order clustering approach to utilize the co-expression dynamics information. This includes two steps: (i) for any
two genes connected with an edge in the differential co-expression network, we calculate the expression correlation in each of the 32 cancer microarray
datasets and store it in a vector, termed the first-order expression correlation profile of the genes; (ii) we then perform hierarchical clustering of all the
gene pairs based on the Euclidean distance between the first-order expression correlation profiles. Unlike commonly used clustering approaches, the unit of the
second-order clustering is a gene pair instead of a gene, and the distance between units is computed based on the first-order expression correlation profiles
instead of the original gene expression profiles, hence the term ``second-order'' clustering ~\cite{zhou2005faa}. Since each edge represents a frequently
occurring co-expression relationship in multiple cancer datasets, it likely represents a functional link. If a cluster of gene pairs follows the same
co-expression pattern across multiple cancer datasets, it represents a module of functional links being turned on or off simultaneously across different cancer
phenotypes.

Given a second-order cluster of gene pairs, we further identify connected network components among them. We suggest that a set of gene pairs is more likely to
be functionally related if they form a connected component. We found that the disjoint network modules of same second-order cluster generally fall into
different functional categories. From the 162 second-order clusters, we measure the functional similarity of genes within each connect network, and between
networks of the same cluster using number of shared Gene Ontology functions. A t-test of the distinction in functional similarities gives $p$-value $1.6 \times
10^{-8}$ on our selected network modules shows that the genes tend to be significantly more functionally coherent within a network than between networks. Thus
each network is more likely to represent a pathway. But in addition, these networks are in the same second-order cluster, this indicates that these networks
are activated together in certain phenotypes.

\begin{figure}[tph]
\centering
\includegraphics [width=\columnwidth, viewport=1 1 1409 978, clip] {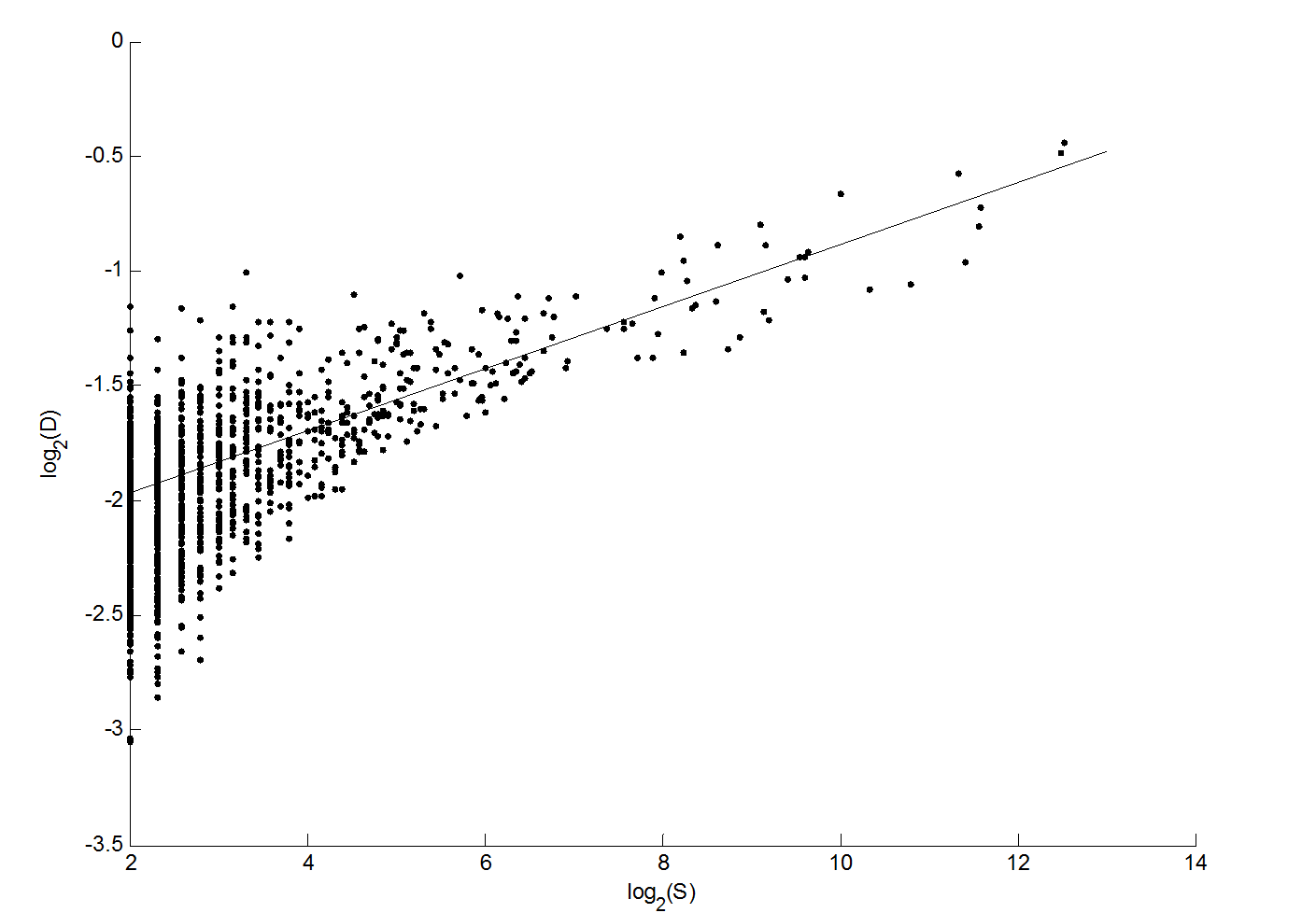}
\caption[The linear relationship between network size $S$ and cluster diameter
$D$.] {The linear relationship between network size $S$ and cluster diameter
$D$. Each dot in the figure corresponds to a connected network of size $\geq$
4. The line is fitted using networks with size $\geq$ 10.}
\label{fig:cancer_scaling}
\end{figure}

Given a second-order hierarchical clustering tree, we traverse the tree bottom up to retrieve connected network components. In general, the size of a connected
component ($S$, the number of edges) decreases with the second-order diameter ($D$), defined as the largest pairwise second-order distance. We found that $S$
and $D$ show a linear scaling relationship in a logarithm scale (Figure \ref{fig:cancer_scaling}). We are especially interested in outliers - network
components small in $D$ but large in $S$, which represent tightly clustered network modules relative to their size. We define the modularity score $\lambda$ of
a subnetwork using a linear scaling model $\lambda = \alpha \log_2(S) - \log_2(D) - \beta$, where $\alpha$ and $\beta$ are estimated using linear fitting. With
our data, we obtain $\alpha = 0.13$ and $\beta = 2.2$. We select the top 60\% of networks ($S \geq 4$) ranked by $\lambda$ scores, removing those networks
having $D \leq 0.34$, and merging heavily overlapping networks. This procedure resulted in 162 second-order clusters comprising 224 connected network modules,
with size ranging from 4 to 64 edges. 
175 (78\%) modules are statistically significantly functionally homogenous based on the GeneOntology Biological Process
annotation (hypergeometric test $p$-value $<$ 0.01). The most predominant functional categories are cell cycle, cell division, cell proliferation, response to
stress, immune response and cell adhesion 
consistent with known pathological mechanisms of cancer.

\begin{figure}[tph]
\centering
\includegraphics [width=0.6\columnwidth] {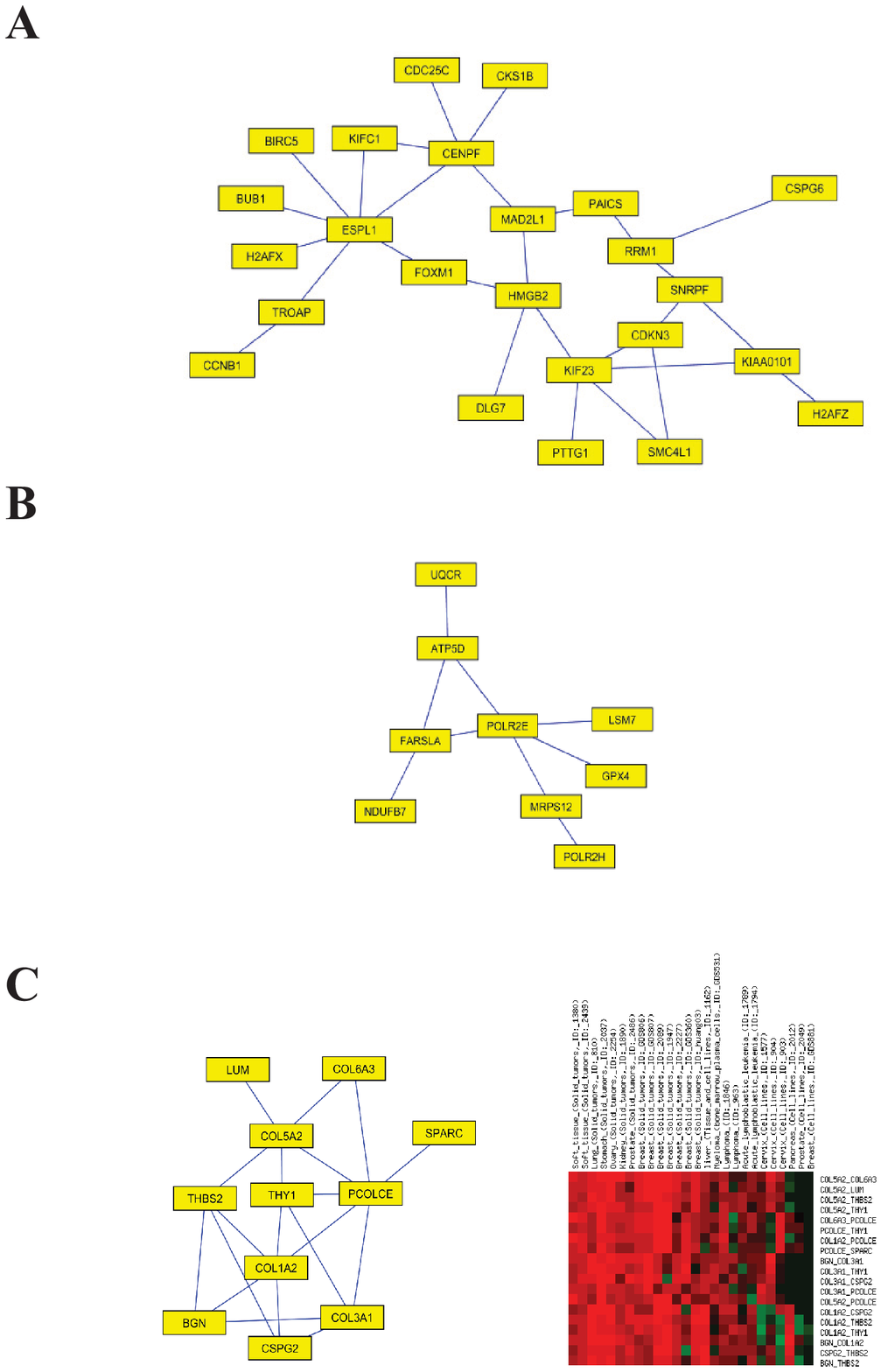}
\caption[General cancer and solid tumor network modules.] {General cancer and solid tumor network modules. (A) A network module that tend to activate in most of cancer datasets,
consisting 24 genes and 28 edges. Average correlation across all data sets is 0.42. Most of genes in the module are related to cell division and genetic stability.(B) Another
network module that is activated in most of cancer datasets, consisting 9 genes and 9 edges. The module is located in the same second order cluster as the one in figure 2a. Its
average correlation across all datasets is 0.39. Most of genes in the module are related to nucleobase, nucleoside, nucleotide and nucleic acid metabolism. (C) Left : a network
that tends to be activated only in solid tumor datasets. Right, the co-expression heatmap of the edges across datasets. Six datasets are not shown in the heatmap due to lack of
valid co-expression estimations. Average correlation in solid tumor datasets and other datasets are 0.61 and 0.17, respectively.} \label{fig:cancer_motif_2}
\end{figure}

One main feature of our approach is that it can simultaneously discover network modules and the types of cancer in which the modules are activated. Figure
~\ref{fig:cancer_motif_2}a shows a module that is activated in most of the cancer datasets. The genes of the module are mostly involved in cell division and genetic stability,
representing a cell proliferation signature, a key feature of cancer. Figure ~\ref{fig:cancer_motif_2}c shows a network module which tends to be activated in only solid tumors.
The genes of the module are mostly involved in cell adhesion and organogenesis, which is specific to solid tumor versus blood cancers or neoplastic cell lines. In the next
section, we will detail two network modules which are activated predominately in breast cancer data sets.

\subsection{Network modules in the breast cancer cluster}
Our analysis resulted in a second-order cluster containing two connected network modules that tend to be more active in all seven breast tumor datasets
relative to the rest of the datasets. The average correlation of these modules in breast tumor and other cancer datasets are 0.49 and 0.23 respectively (the
t-test of co-expressions between breast tumor and the rest of cancer datasets gives a $p$-value of $1.56 \times 10 ^{-95}$).

\subsubsection[A tumor suppressor network related to PDGF superfamily signaling]{A tumor suppressor network related to PDGF superfamily \\ signaling}
\begin{figure}[tph]
\centering
\includegraphics [width=0.7\columnwidth] {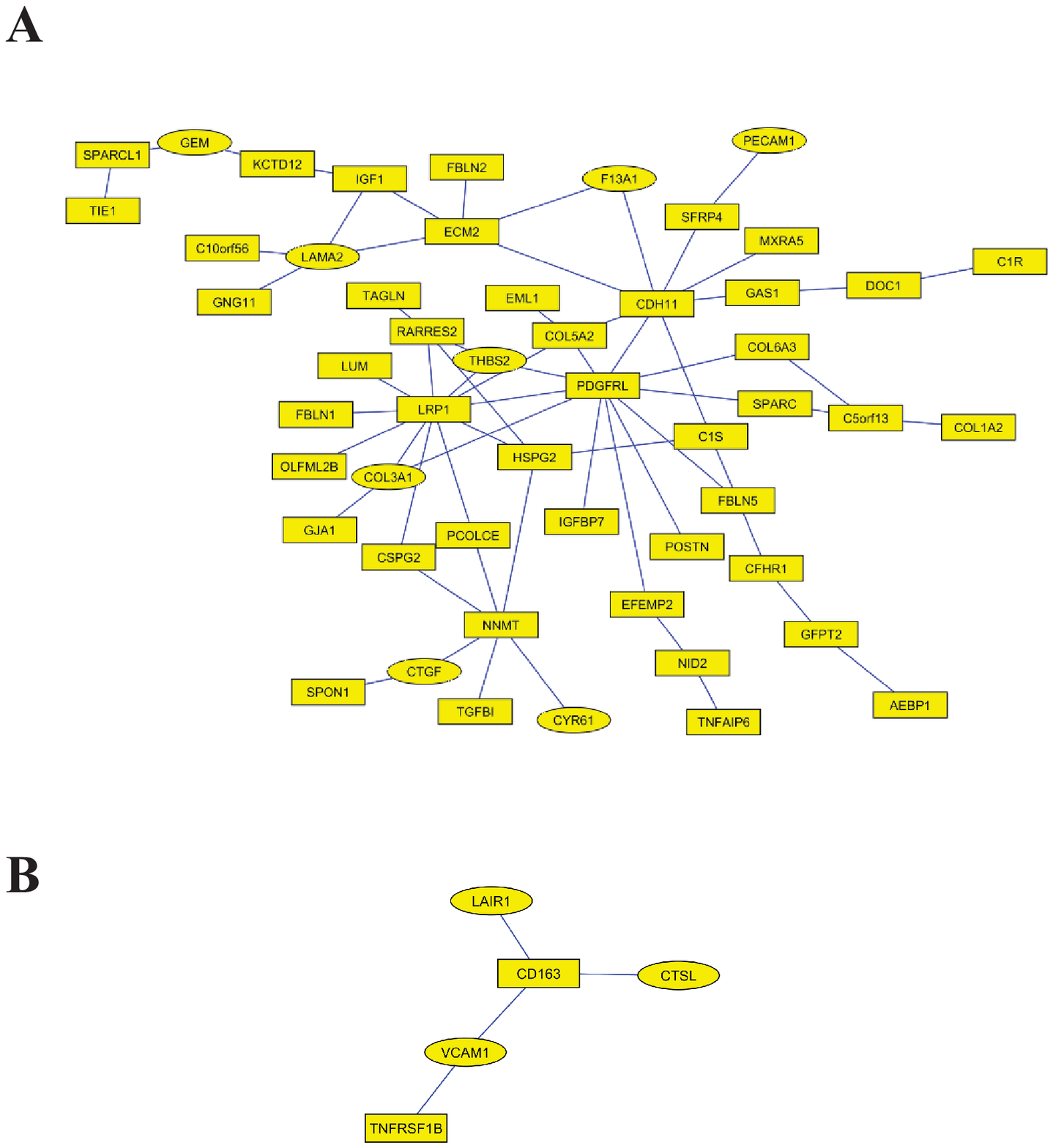}
\caption[Coordinated breast tumor network modules.]{Coordinated breast tumor network modules. (A) A breast tumor network module that involved in PDGF signalling. Genes in round
shape have promoter regions that are predicted to be bound by transcription factor ZNF148. (B) A breast tumor network module related to inflammatory response, located in the same
second-order cluster as the one in Figure 3a. Genes in round shape have promoter regions that are predicted to be bound by transcription factor POU2F1.} \label{fig:cancer_motif_3}
\end{figure}

The module in Figure ~\ref{fig:cancer_motif_3}a contains 52 genes. Most of them are extracellular
or membrane proteins, and 23 genes have previously been found to be involved in breast cancer.
Following are examples of genes in the breast tumor suppressor modules that known to related to
breast cancer : connective tissue growth factor (CTGF) is over-expressed in breast cancer cell
lines with increased metastatic activity \cite{kang2003mpm}; Fibulin 1 (FBLN1) is implicated in
immune response against breast cancer \cite{pupa2003iap}, and one of its splice variants is
over-expressed in breast \cite{bardin2005tap}; the cysteine-rich secreted protein (SPARC) plays a
crucial role in tumor development in breast cancer \cite{watkins2005ils}; GAS1 is also found to be
induced in apoptotic mammary gland cells \cite{seol2005epa}; Cysteine-rich angiogenic inducer 61
(CYR61) is involved in the proliferation, cell survival, and Taxol resistance of breast cancer
\cite{menendez2004nct}; finally, in the case of a hub gene LRP1 (low density lipoprotein
receptor-related protein 1), the T allele of the C766T polymorphism is associated with an increased
risk of breast cancer development \cite{benevs2003cld}.

Among the 52 genes, 16 are involved in cell adhesion ($p$-value $7.4 \times 10 ^{-11}$ ), and 14 are involved cell-cell signalling ($p$-value $7.9 \times 10 ^{-6}$ ), suggesting a
role in tumor invasiveness of the module ~\cite{ross2004bcb}.

General cancer and solid tumor network modules Most interestingly, we found one main function of this network module appears to be tumor suppression via the
inhibition of PDGF superfamily signaling. A hub gene with high degree of this module is the gene PDGF receptor-like (PDGFRL) (degree 11). While its precise
biological function is not known, PDGFRL encodes a 375aa product with significant sequence similarity to the extracellular domain of PDGFR. Indeed, mutations
in PDGFRL have been found in individual cancer samples ~\cite{komiya1997pga, lerebours1999fdm, an2002dts, kahng2003lhc}. PDGFRL is located in chromosome
8p22-8p21.3, where multiple studies have suggested the existence of a putative breast cancer tumor suppressor gene ~\cite{yaremko1996lhs, seitz2000rdm,
rennstam2003pci}. Recently, an in-depth study of the region using microcell-mediated chromosome transfer found that indeed PDGFRL expression is decreased in
the majority of breast cancer cells ~\cite{seitz2006gbd}. Many genes in this network module have been found to be involved in PDGF superfamily signaling. For
example, Cysteine-rich protein (SPARC) binds to PDGF-AB and PDGF-BB dimers, inhibiting the binding of these growth factors to their cell surface receptors
~\cite{raines1992egs} and inhibiting PDGF-induced vascular smooth muscle proliferation ~\cite{motamed2002ips}.  Also, connective tissue growth factor (CTGF)
has structural similarities with PDGF ~\cite{kanazawa2005epd}. Recently, two new PDGF ligands were discovered that have a N-terminal complement subcomponent
C1R/C1S, UEGf, BMP1 (CUB) domain ~\cite{ustach2005pdg}. Interestingly both C1R and C1S are members of this network module. Also, LRP1 is a physiological
modulator of the PDGF signaling pathway ~\cite{takayama2005ldl}. In total, we found direct Pubmed literature support for 19 out of the 52 member genes to be
involved in (or related to) the PDGF-superfamily signalling. Furthermore, it is known that zinc finger protein (ZNF148) binds to the PDGFR
~\cite{debustos2005ppp} gene promoter. We screened the promoter regions of the 52 member genes, and found that the binding sites of ZNF148 are significantly
enriched (hypergeometric $p$-value 0.016). In addition, it has been reported that PDGFR positively regulates collagen production (for example
~\cite{chaudhary2007ipv, zymek2006rpd}). Our module showed the breast tumor specific co-expression between PDGFRL and collagens COL3A1, COL5A2 and COL6A3. Many
of the above evidences support the hypothesis that PDGFRL has an agonistic function to PDGFR signaling.

Although abundant evidence suggests the individual involvement of many of these genes in tumor suppression, their coordinated function has never been elucidated. Here by adopting
a network perspective, we have identified their interrelationships and a gene (PDGFRL) that may play a central role in this tumor suppressor network.

\subsubsection{A network module related to inflammatory response}

Another identified breast cancer-specific network module (Figure ~\ref{fig:cancer_motif_3}b) may be involved in the coordination of the inflammatory response to cancer pathology.
This network consists of 5 genes: tumor necrosis factor receptor superfamily member 1B (TNFRSF1B); vascular cell adhesion molecule 1 (VCAM1); leukocyte-associated
immunoglobulin-like receptor 1 (LAIR1); and Cathepsin L1 (CSTL). They are arranged around the macrophage-associated antigen (CD163). Several lines of evidence have implicated
these genes individually in breast cancer, for example: increased plasma levels of VCAM1 is associated with advanced breast cancer ~\cite{silva2006sva}; genetic variation in
TNFRSF1B may predict the late onset of breast carcinoma, and relapse and death for patients with breast carcinoma ~\cite{mestiri2005fpt}; finally, the breast cancer cell line
exhibiting the highest in vitro invasiveness also expressed the highest amount of CTSL1 splice variant L-A3 ~\cite{caserman2006csv}.

Most of the genes of this module are related to tumor necrosis factor (TNF), an inflammatory cytokine. It has been reported that activation of rat CD163 on peritoneal macrophages
induces the production of pro-inflammatory mediators including TNF \\
\cite{polfliet2006rms}; TNF directly interacts with TNFRSF1B ~\cite{medvedev1996drt} and is a mediator of TNF
function in the mouse ovary ~\cite{greenfeld2007tnf}. Finally, transcription of VCAM1 in endothelial cells can be induced by TNF ~\cite{neish1995eir}.

A transcription factor, octamer-binding transcription factor-1 (POU2F1, also known as Oct-1) has been predicted to bind promoter region of VCAM1, LAIR1 and CTSL (hypergeometric
$p$-value 0.012). The binding of POU2F1 to VCAM1 promoter is indeed supported by the literature ~\cite{schwachtgen1998oit, syder2003ipc, ortego2002hmc}. Also, POU2F1 has been
found to bind SP3 ~\cite{liu2003gtf}, which is reported to activate the transcription of CSTL ~\cite{sriraman2004cge}.

The tight and coordinated expression of the genes in this network module reveals an induced inflammatory response that may be important in breast tumor progression
~\cite{coussens2004iac}.

\subsection[Identification of pathway coordination beyond co-expression]{Identification of pathway coordination beyond co-expression}
A major advantage of second-order clustering is that it can identify functionally related genes
beyond co-expression, as illustrated in Figure ~\ref{fig:cancer_motif_1}b. We elaborate on this
point in this section. To allow readers to easily assess the magnitude of the correlation, only in
this section we use the Pearson correlation to measure the co-expression level.

Based on the definition of second-order clustering, connected network components within the same second-order cluster show coordinated activities, which implies their functional
relevance. In the example of the two modules in the general cancer cluster in Figure ~\ref{fig:cancer_motif_2}a and 2b, each module may play different roles in the regulation of
specific biological processes -- cell division and nucleic acid metabolism, respectively; the latter is clearly required for cell division. Given the fact that these processes
belong to the same second-order cluster, they may represent facets of the same underlying neoplastic process. However, member genes of the two modules exhibit distinct expression
patterns: The average Pearson correlation between genes of the two modules is only 0.13.

As another example, the two breast cancer modules described in last section are also related. Indeed, the collaboration of PDGF signalling and TNF have long been known to be
required for tissue repairing ~\cite{ottaviani2004ihf}, and their abnormal expression play important but partially defined roles in breast tumor development and progression
~\cite{chala2006irg}. On the other hand, member genes of the two modules exhibit relatively distinct expression patterns. For example, two hub genes of modules, PDGFRL and CD163,
also show very weak expression similarities across all seven breast cancer datasets:  the average Pearson correlation between these two genes is only 0.24.

Besides the above examples, we found coordinated modules within the majority of identified second-order clusters. Overall, from the total 162 second-order clusters, 25\% give rise to more than one connected network module. To estimate the amount of cross module co-expression within second-order clusters, for each second order cluster, we first determined the active cancer datasets, in which the average Pearson correlation of gene pairs in the cluster is greater than 0.5. For 72\% of those module pairs within the same second-order cluster, the average gene pairwise Pearson correlation between modules in the active datasets is less than 0.5  (normalized Percentage Bend correlation approximately $<$ 0.35), and for 30\% module pairs the cross-module average gene pairwise Pearson correlation is less than 0.3 (normalized Percentage Bend correlation approximately $<$ 0.19).

Furthermore, even genes in the same network module are not necessarily highly co-expressed when the module is active, despite their high degree of functional homogeneity, as discussed previously. We found that in 32 of 224 (14 \%) modules, the average pairwise Pearson correlation of any two genes in the module is $<$ 0.5 in the corresponding active datasets. Such modules could therefore easily be overlooked by traditional clustering methods.

\section{Discussion}
The rapid accumulation of microarray data provides unprecedented opportunities to study the molecular mechanisms underlying disease pathogenesis and progression. Although many
studies utilized multiple microarray datasets to derive consistent lists of genes specific for (subtypes of) cancer ~\cite{rhodes2002mam, rhodes2004lsm, kyoonchoi2004iam}, little
attention has been paid to derive genetic networks characterizing different types of cancer. Segal et al. ~\cite{segal2004mms} used predefined biologically meaningful gene sets
including known biological pathways, and have successfully identified activated or repressed biological modules in a wide variety of neoplastic conditions. The approach, however,
relies on the knowledge of pre-defined biological modules and has limited use in the discovery of novel association between genes. A recent study by Choi et al.
~\cite{choi2005dca} compared the two co-expression networks summarized from 10 tumor and normal datasets, respectively, and have identified functional differences between normal
growth and cancer in terms of gene coexpression changes in broad areas of physiology. However, due to the multifaceted nature of cancer, interactions in such a derived summary
network may not be simultaneously active in individual datasets, i.e. specific cancer conditions.

In this study, we propose an unsupervised method that integrates both co-expression dynamics and network topology information to characterize cancer (subtype) specific network
modules. The identified modules, such as modules activated across all cancer subtypes or only in solid tumors, are novel, but consistent with known molecular mechanisms.
Importantly, we have discovered a potential tumor suppressor network particularly active in breast tumors, and provide compelling evidence that the hub gene PDGFRL is a true tumor
suppressor gene. Compared to commonly used differential or co-expression analysis, our approach has the following advantages: (1) our unsupervised approach simultaneously
discovers network modules and the conditions (e.g. cancer subtypes) in which they are activated, thus providing new insights into the complex cellular mechanisms that characterize
cancer and cancer subtypes.  (2) Compared to existing approaches ~\cite{choi2005dca, liu2006tbc, carter2004gce, steuer2003oai, butte2000dfr} which can only identify densely
connected network modules based solely on network topology information, our  approach incorporating co-expression dynamics information (second-order similarity) can extract more
diverse types of modules regardless of network density. It is known that many biological pathways do not necessarily form densely connected modules. (3) Our approach can reveal
coordination of pathways beyond co-expression. (4) Our method can be applied to any types of molecular networks beyond co-expression network, when data of multiple networks under
different conditions are available.

In the current framework, the selection of biologically meaningful modules still need certain amount of manual intervention from biology expert. We are looking for more systematic
ways for module selection, especially by putting the framework into the context of network statistics to improve the robustness. For example, although the scaling model we
constructed is based on direct observation of the distribution of network properties S and D, their log-linear relationship suggests that it should be straightforward to make use
of the exponential random graph models that have been used in recent years to study statistical aspects of networks ~\cite{carrington2005mam}. In essence, such models linearly
combine network properties and assign the probability of observed networks as the exponential of such linear combinations. Integrating this work with the exponential family
probabilistic models may provide both better estimation of the coefficients in the linear scaling model and more accurate selection of network modules via hypothesis testing. We
intend to explore this direction in future work.

The choice of datasets depends on the research question. Ideally there should be a balanced and sufficient sampling of different phenotypes (in particular different tissues for
this cancer study). Particularly, a paired Wilcoxon test between cancer and normal samples of the same tissue would significantly eliminate the amount of tissue-specific
co-expressions. However, due to the limited amount and the heterogeneity of existing data it is currently impractical to achieve this goal. Using a weighted sampling scheme could
potentially bypass the imbalance effects. We aim to investigate this strategy, and use statistical models of the correlation values to determine the weighting factors.

\section{Methods}
\subsection{Datasets}
We curated 32 cancer and 23 non-cancer human gene expression datasets mainly from the Stanford Microarray Database (SMD) and Gene Expression Omnibus (GEO) databases, each
containing more than 15 microarrays, on either Affymetrix or cDNA platforms. In each dataset, if there are multiple probes that correspond to the same gene, we choose the one that
contains the least amount of missing values. For datasets containing absolute expression measurements, we convert all values $\leq$ 10 to 10, then perform a base 2 log transform.
Estimation of Pairwise gene co-expression We used Percentage Bend Correlation ~\cite{wilcox2005ire} (with $\beta=0.05$) to obtain a robust correlation estimate. Percentage Bend
Correlation first detects outliers in expression values of each gene then reduces the effects of those outliers in the correlation calculation. Only gene pairs with a large number
of valid samples $m \geq 15$ are used to calculate correlation. To make the correlation estimates comparable across different datasets of variable sample sizes and among different
gene pairs of different amount of missing expression measures within the same dataset, we performed Fisher's z-transform ~\cite{anderson:ims} to reduce sample size effect. Given a
correlation estimate r and sample size n, the Fisher's Z scores (divided by its theoretical standard deviation) is calculated as $z = \frac{\sqrt{n-3}}{2} \log{( \frac{1+r}{1-r}
)}$, which theoretically has an asymptotically standard normal distribution. Note that sample size $n$ may be different from gene pair to gene pair due to missing values, and from
dataset to dataset. In reality, we observed that the distributions of the z-score are still different from dataset to dataset: we therefore normalized z-scores to enforce the
standard normal distribution. After that, standardized correlations $r'$ are obtained by inverting the z-score with a fixed $n$ of 30.

\subsection{Select differentially co-expressed gene pairs}
We define a gene pair to be differentially co-expressed between cancer and non-cancer if it satisfies the following two criteria: (1) their expression correlation in cancer
datasets is sufficiently strong (can be either positively or negatively high). This is done by setting threshold for average summed square of correlations in cancer datasets. i.e,
$\sqrt{\frac{1}{c} \sum_k{(r_i^{(k)})^2}} < 0.35$  for the gene pair $i$, where there are $c$ valid correlation estimations ($c=32$ if there are no missing values) and $k$ is
dataset index corresponding to all valid correlation estimations; and (2) the Wilcoxon ranksum test of correlations between cancer and non-cancer datasets gives a $p$-value $\leq$
0.01.

\subsection[Identify conditionally activated network module candidates]{Identify conditionally activated network module candidates}
We hierarchically clustered the differentially co-expressed gene pairs based on their expression
correlation profiles using the CLUSTER program ~\cite{dehoon2004osc} with complete linkage and
Euclidean distance. The Euclidean distance is averaged to provide a simple estimation given the
existence of missing correlations (due to the missing value problem). $d_{i,j} = \sqrt{\frac{1}{c}
\sum_k{(r_i^{(k)} - r_j^{(k)})}}$ where $r_i^{(k)}$ and $r_j^{(k)}$  are the correlations of gene
pairs $i$ and $j$ in the dataset $k$, respectively, and $c$ is number of valid correlations. In the
hierarchical tree, each leaf node represents a gene pair, and each inner node corresponds to a
second-order cluster of gene pairs (edges) which may comprise zero, one, or more connected network
components. In cases where the size of the differential network is too big to be processed using
hierarchical clustering (HC), the gene pairs were first separated using k-means clustering then
processed the smaller clusters separately by hierarchical clustering. As for our experience, the
biologically meaningful modules normally contain less than a few hundred edges, thus k-means
clustering will keep most of modules intact.

\subsection[Gene ontology function and transcription factor enrichment of modules]{Gene ontology function and transcription factor enrichment of modules}
The functional enrichment analysis is done by the hypergeometric test on genes. We selected 419
Gene Ontology (GO) functions (i.e. biological process terms) which are 4 levels below the root in
the GO hierarchy. Each gene may be directly or indirectly associated with some of these functions.
A set of genes will be considered to have a enriched function when (1) the functional homogeneity
modeled by the hypergeometric distribution ~\cite{wu2002lsp} is significant at a significance level
0.01 and (2) there are at least 2 genes in the set are associated with the function.

\subsection{Identification of transcription factor binding}
10kb upstream sequences for each gene were obtained from NCBI Gene database. After applying RepeatMasker ~\cite{smit2004ro}, we used the MATCH program of TRANSFAC
~\cite{matys2003trt} (version 9.2) to scan the sequences for the presence of transcription factor binding sites based on position weight matrices. We used vertebrate-specific
matrices, and chose cut-offs to minimize the sum of false positives and false negatives. We kept only the top 3,000 hits per matrix, sorted by the matrix similarity score.
Altogether we obtained 349,178 predicted transcription factor target relationships of 180 transcription factors. A hypergeometric test was performed for each network module to
search for over-represented transcription factor binding sites.

\begin{singlespace}
\references[References]{alpha}{reference}

\end{singlespace}


\newcommand{\etalchar}[1]{$^{#1}$}
\begin{thebibliography}{KCYCGK{\etalchar{+}}04}

\bibitem[ABN{\etalchar{+}}99]{alon1999bpg}
U.~Alon, N.~Barkai, DA~Notterman, K.~Gish, S.~Ybarra, D.~Mack, and AJ~Levine.
\newblock {Broad patterns of gene expression revealed by clustering analysis of
  tumor and normal colon tissues probed by oligonucleotide arrays}, 1999.

\bibitem[ABO{\etalchar{+}}00]{albitar2000dbr}
M.~Albitar, M.~Beran, S.~O'Brien, H.~Kantarjian, E.~Frieriech, M.~Keating, and
  E.~Estey.
\newblock {Differences between refractory anemia with excess blasts in
  transformation and acute myeloid leukemia}.
\newblock {\em Blood}, 96(1):372, 2000.

\bibitem[ALG{\etalchar{+}}02]{an2002dts}
Q.~An, Y.~Liu, Y.~Gao, J.~Huang, X.~Fong, L.~Liu, D.~Zhang, J.~Zhang, and
  S.~Cheng.
\newblock {Deletion of tumor suppressor genes in Chinese non-small cell lung
  cancer.}
\newblock {\em Cancer letters}, 184(2):189, 2002.

\bibitem[AM02]{ambroise2002sbg}
C.~Ambroise and G.J. McLachlan.
\newblock {Selection bias in gene extraction on the basis of microarray
  gene-expression data}.
\newblock {\em Proceedings of the National Academy of Sciences}, 99(10):6562,
  2002.

\bibitem[And]{anderson:ims}
TW~Anderson.
\newblock {An Introduction to Multivariate Statistical Analysis. 2003}.

\bibitem[Aro01]{aronson2001emb}
A.R. Aronson.
\newblock {Effective mapping of biomedical text to the UMLS Metathesaurus: the
  MetaMap program}.
\newblock {\em Proc AMIA Symp}, 17(21):17--21, 2001.

\bibitem[BBG{\etalchar{+}}05]{balog2005cit}
A.~Balog, Z.~Borb{\'e}nyi, Z.~Gyulai, L.~Molnar, and Y.~Mandi.
\newblock {Clinical Importance of Transforming Growth Factor-b but Not of Tumor
  Necrosis Factor-a Gene Polymorphisms in Patients with the Myelodysplastic
  Syndrome Belonging to the Refractory Anemia Subtype}.
\newblock {\em Pathobiology}, 72(3):165, 2005.

\bibitem[BJ{\v{Z}}{\etalchar{+}}03]{benevs2003cld}
P.~Bene{\v{s}}, M.~Jurajda, J.~{\v{Z}}aloud{\'\i}k,
  L.~Izakovi{\v{c}}ov{\'a}-Holl{\'a}, and J.~V{\'a}cha.
\newblock {C766T low-density lipoprotein receptor-related protein 1 (LRP1) gene
  polymorphism and susceptibility to breast cancer}.
\newblock {\em Breast Cancer Res}, 5(3):R77--R81, 2003.

\bibitem[BK06]{butte2006cai}
A.J. Butte and I.S. Kohane.
\newblock {Creation and implications of a phenome-genome network}.
\newblock {\em Nature biotechnology}, 24:55--62, 2006.

\bibitem[BMC{\etalchar{+}}00]{bittner2000mcc}
M.~Bittner, P.~Meltzer, Y.~Chen, Y.~Jiang, E.~Seftor, M.~Hendrix, M.~Radmacher,
  R.~Simon, Z.~Yakhini, A.~Ben-Dor, et~al.
\newblock {Molecular classification of cutaneous malignant melanoma by gene
  expression profiling}.
\newblock {\em Nature}, 406(6795):536--540, 2000.

\bibitem[BMM{\etalchar{+}}05]{bardin2005tap}
A.~Bardin, F.~Moll, R.~Margueron, C.~Delfour, ML~Chu, T.~Maudelonde,
  V.~Cavailles, and P.~Pujol.
\newblock {Transcriptional and posttranscriptional regulation of fibulin-1 by
  estrogens leads to differential induction of messenger ribonucleic acid
  variants in ovarian and breast cancer cells}.
\newblock {\em Endocrinology}, 146(2):760--768, 2005.

\bibitem[Bod]{bodenreider32uml}
O.~Bodenreider.
\newblock {The Unified Medical Language System (UMLS): integrating biomedical
  terminology}.
\newblock {\em Nucleic Acids Research}, 32(90001):267--270.

\bibitem[BSS93]{bazaraa1993}
{Mokhtar S.} Bazaraa, {Hanif D.} Sherali, and {C. M.} Shetty.
\newblock {\em Nonlinear programming : theory and algorithms}.
\newblock Wiley, 2. ed edition, 1993.

\bibitem[BTGM04]{brunet2004mam}
J.P. Brunet, P.~Tamayo, T.R. Golub, and J.P. Mesirov.
\newblock {Metagenes and molecular pattern discovery using matrix
  factorization}.
\newblock {\em Proceedings of the National Academy of Sciences},
  101(12):4164--4169, 2004.

\bibitem[BTS{\etalchar{+}}00]{butte2000dfr}
A.J. Butte, P.~Tamayo, D.~Slonim, T.R. Golub, and I.S. Kohane.
\newblock {Discovering functional relationships between RNA expression and
  chemotherapeutic susceptibility using relevance networks}.
\newblock {\em Proceedings of the National Academy of Sciences}, 97(22):12182,
  2000.

\bibitem[BTW{\etalchar{+}}07]{barrett2007ngm}
T.~Barrett, D.B. Troup, S.E. Wilhite, P.~Ledoux, D.~Rudnev, C.~Evangelista,
  I.F. Kim, A.~Soboleva, M.~Tomashevsky, and R.~Edgar.
\newblock {NCBI GEO: mining tens of millions of expression profiles--database
  and tools update}.
\newblock {\em Nucleic Acids Research}, 35(Database issue):D760, 2007.

\bibitem[BvD04]{brunner2004sff}
HG~Brunner and MA~van Driel.
\newblock {From syndrome families to functional genomics.}
\newblock {\em Nat Rev Genet}, 5(7):545--51, 2004.

\bibitem[CBGB04]{carter2004gce}
S.L. Carter, C.M. Brechbuhler, M.~Griffin, and A.T. Bond.
\newblock {Gene co-expression network topology provides a framework for
  molecular characterization of cellular state}, 2004.

\bibitem[CCS{\etalchar{+}}02]{chen2002gep}
X.~Chen, S.T. Cheung, S.~So, S.T. Fan, C.~Barry, J.~Higgins, K.M. Lai, J.~Ji,
  S.~Dudoit, I.O.L. Ng, et~al.
\newblock {Gene expression patterns in human liver cancers}.
\newblock {\em Gene Expression Patterns}, 13(6):1929--1939, 2002.

\bibitem[CKSL06]{caserman2006csv}
S.~Caserman, S.~Kenig, B.F. Sloane, and T.T. Lah.
\newblock {Cathepsin L splice variants in human breast cell lines.}
\newblock {\em Biological Chemistry}, 387(5):629, 2006.

\bibitem[CMI{\etalchar{+}}06]{chala2006irg}
E.~Chala, C.~Manes, H.~Iliades, G.~Skaragkas, D.~Mouratidou, and E.~Kapantais.
\newblock {Insulin resistance, growth factors and cytokine levels in overweight
  women with breast cancer before and after chemotherapy}.
\newblock {\em Hormones}, 5(2):137--146, 2006.

\bibitem[Cou02]{coussens2004iac}
Z~Coussens, L.and~Werb.
\newblock {Inflammation and cancer}.
\newblock {\em Nature}, 420(6917):860--867, 2002.

\bibitem[Cov65]{cover1965gas}
T.M. Cover.
\newblock {Geometrical and statistical properties of systems of linear
  inequalities with applications in pattern recognition}.
\newblock {\em IEEE transactions on electronic computers}, 14(3):326--334,
  1965.

\bibitem[CRH{\etalchar{+}}07]{chaudhary2007ipv}
NI~Chaudhary, GJ~Roth, F.~Hilberg, J.~Muller-Quernheim, A.~Prasse, G.~Zissel,
  A.~Schnapp, and JE~Park.
\newblock {Inhibition of PDGF, VEGF and FGF signalling attenuates fibrosis}.
\newblock {\em European Respiratory Journal}, 29(5):976, 2007.

\bibitem[CSW05]{carrington2005mam}
P.J. Carrington, J.~Scott, and S.~Wasserman.
\newblock {\em {Models and methods in social network analysis}}.
\newblock Cambridge University Press, 2005.

\bibitem[CYYK05]{choi2005dca}
J.K. Choi, U.~Yu, O.J. Yoo, and S.~Kim.
\newblock {Differential coexpression analysis using microarray data and its
  application to human cancer}.
\newblock {\em Bioinformatics}, 21(24):4348--4355, 2005.

\bibitem[DB02]{dettling2002scg}
M.~Dettling and P.~Buhlmann.
\newblock {Supervised clustering of genes}.
\newblock {\em Genome Biology}, 3(12):1--0069, 2002.

\bibitem[DBSS{\etalchar{+}}05]{debustos2005ppp}
C.~De~Bustos, A.~Smits, B.~Stromberg, VP~Collins, M.~Nister, and G.~Afink.
\newblock {A PDGFRA promoter polymorphism, which disrupts the binding of
  ZNF148, is associated with primitive neuroectodermal tumours and
  ependymomas}, 2005.

\bibitem[DD03]{deininger2003stt}
M.W.N. Deininger and B.J. Druker.
\newblock {Specific Targeted Therapy of Chronic Myelogenous Leukemia with
  Imatinib}.
\newblock {\em Pharmacological Reviews}, 55(3):401--423, 2003.

\bibitem[dHINM04]{dehoon2004osc}
M.J.L. de~Hoon, S.~Imoto, J.~Nolan, and S.~Miyano.
\newblock {Open source clustering software}.
\newblock {\em Bioinformatics}, 20(9):1453, 2004.

\bibitem[DLCBPS{\etalchar{+}}08]{delourdeschauffaille2008apl}
M.~De~Lourdes~Chauffaille, D.~Borri, R.~Proto-Siqueira, ES~Moreira, and
  FL~Alberto.
\newblock {Acute promyelocytic leukemia with t (15; 17): frequency of
  additional clonal chromosome abnormalities and FLT3 mutations.}
\newblock {\em Leukemia \& lymphoma}, 49(12):2387, 2008.

\bibitem[FF08]{fan2008hdc}
J.~Fan and Y.~Fan.
\newblock {High dimensional classification using features annealed independence
  rules}.
\newblock {\em Annals of statistics}, 36(6):2605, 2008.

\bibitem[FS03]{freimer2003hpp}
N.~Freimer and C.~Sabatti.
\newblock {The Human Phenome Project}.
\newblock {\em Nature Genetics}, 34(1):15--21, 2003.

\bibitem[FSMJ03]{furlanello2003ebg}
C.~Furlanello, M.~Serafini, S.~Merler, and G.~Jurman.
\newblock {Entropy-based gene ranking without selection bias for the predictive
  classification of microarray data}.
\newblock {\em BMC bioinformatics}, 4(1):54, 2003.

\bibitem[Gen]{gen14ecm}
G.Z.M. Gen.
\newblock {Evolutionary Computation on Multicriteria Production Process
  Planning Problem}.
\newblock In {\em Proceedings of the... IEEE Conference on Evolutionary
  Computation}, volume~14, page 419. IEEE.

\bibitem[GRP{\etalchar{+}}07]{greenfeld2007tnf}
C.R. Greenfeld, K.F. Roby, M.E. Pepling, J.K. Babus, P.F. Terranova, and J.A.
  Flaws.
\newblock {Tumor necrosis factor (TNF) receptor type 2 is an important mediator
  of TNF alpha function in the mouse ovary}.
\newblock {\em Biology of reproduction}, 76(2):224, 2007.

\bibitem[GST{\etalchar{+}}99]{golub1999mcc}
TR~Golub, DK~Slonim, P.~Tamayo, C.~Huard, M.~Gaasenbeek, JP~Mesirov, H.~Coller,
  ML~Loh, JR~Downing, MA~Caligiuri, et~al.
\newblock {Molecular classification of cancer: class discovery and class
  prediction by gene expression monitoring}.
\newblock {\em Science}, 286(5439):531, 1999.

\bibitem[GWBV02]{guyon2002gsc}
I.~Guyon, J.~Weston, S.~Barnhill, and V.~Vapnik.
\newblock {Gene selection for cancer classification using support vector
  machines}.
\newblock {\em Machine learning}, 46(1):389--422, 2002.

\bibitem[JMRWK04]{jordan2004cac}
I.K. Jordan, L.~Marino-Ramirez, Y.I. Wolf, and E.V. Koonin.
\newblock {Conservation and coevolution in the scale-free human gene
  coexpression network}.
\newblock {\em Molecular biology and evolution}, 21(11):2058--2070, 2004.

\bibitem[JSOP01]{jacob2001res}
MP~Jacob, M.~Sauvage, and M.~Osborne-Pellegrin.
\newblock {Regulation of elastin synthesis}.
\newblock {\em Journal de la Soci{\'e}t{\'e} de biologie}, 195(2):131, 2001.

\bibitem[JSR02]{jager2002igs}
J.~J{\"a}ger, R.~Sengupta, and W.L. Ruzzo.
\newblock {Improved gene selection for classification of microarrays}.
\newblock {\em Biocomputing 2003}, page~53, 2002.

\bibitem[JY03]{jornsten2003sgc}
R.~Jornsten and B.~Yu.
\newblock {Simultaneous gene clustering and subset selection for sample
  classification via MDL}, 2003.

\bibitem[KCYCGK{\etalchar{+}}04]{kyoonchoi2004iam}
J.~Kyoon~Choi, J.~Young~Choi, D.~Ghon~Kim, D.~Wook~Choi, B.~Yeo~Kim, K.~Ho~Lee,
  Y.~Il~Yeom, H.~Sook~Yoo, O.~Joon~Yoo, and S.~Kim.
\newblock {Integrative analysis of multiple gene expression profiles applied to
  liver cancer study}.
\newblock {\em FEBS letters}, 565(1-3):93--100, 2004.

\bibitem[KLK{\etalchar{+}}03]{kahng2003lhc}
Y.S. Kahng, Y.S. Lee, B.K. Kim, W.S. Park, J.Y. Lee, and C.S. Kang.
\newblock {Loss of heterozygosity of chromosome 8p and 11p in the dysplastic
  nodule and hepatocellular carcinoma.}
\newblock {\em Journal of Gastroenterology \& Hepatology}, 18(4):430, 2003.

\bibitem[KMK{\etalchar{+}}05]{kanazawa2005epd}
S.~Kanazawa, T.~Miyake, T.~Kakinuma, K.~Tanemoto, T.~Tsunoda, and K.~Kikuchi.
\newblock {The expression of platelet-derived growth factor and connective
  tissue growth factor in different types of abdominal aortic aneurysms.}
\newblock {\em The Journal of cardiovascular surgery}, 46(3):271, 2005.

\bibitem[KR90]{kaufman1990fgd}
L.~Kaufman and P.J. Rousseeuw.
\newblock {Finding groups in data: an introduction to cluster analysis}.
\newblock {\em New York}, 1990.

\bibitem[KSS{\etalchar{+}}03]{kang2003mpm}
Y.~Kang, P.M. Siegel, W.~Shu, M.~Drobnjak, S.M. Kakonen, C.~Cord{\'o}n-Cardo,
  T.A. Guise, and J.~Massagu{\'e}.
\newblock {A multigenic program mediating breast cancer metastasis to bone}.
\newblock {\em Cancer Cell}, 3(6):537--549, 2003.

\bibitem[KSU{\etalchar{+}}97]{komiya1997pga}
A.~Komiya, H.~Suzuki, T.~Ueda, S.~Aida, N.~Ito, T.~Shiraishi, R.~Yatani,
  M.~Emi, K.~Yasuda, J.~Shimazaki, et~al.
\newblock {PRLTS gene alterations in human prostate cancer}.
\newblock {\em Cancer Science}, 88(4):389--393, 1997.

\bibitem[KWR{\etalchar{+}}01]{khan2001cad}
J.~Khan, J.S. Wei, M.~Ringn{\'e}r, L.H. Saal, M.~Ladanyi, F.~Westermann,
  F.~Berthold, M.~Schwab, C.R. Antonescu, C.~Peterson, et~al.
\newblock {Classification and diagnostic prediction of cancers using gene
  expression profiling and artificial neural networks}.
\newblock {\em Nature medicine}, 7(6):673--679, 2001.

\bibitem[LCL{\etalchar{+}}06]{liu2006tbc}
C.C. Liu, W.S.E. Chen, C.C. Lin, H.C. Liu, H.Y. Chen, P.C. Yang, P.C. Chang,
  and J.J.W. Chen.
\newblock {Topology-based cancer classification and related pathway mining
  using microarray data}.
\newblock {\em Nucleic acids research}, 34(14):4069, 2006.

\bibitem[LKS{\etalchar{+}}07]{lage2007hpi}
K.~Lage, E.O. Karlberg, Z.M. Storling, P.{\'I}. {\'O}lason, A.G. Pedersen,
  O.~Rigina, A.M. Hinsby, Z.~T{\"u}mer, F.~Pociot, N.~Tommerup, et~al.
\newblock {A human phenome-interactome network of protein complexes implicated
  in genetic disorders}.
\newblock {\em Nature Biotechnology}, 25:309--316, 2007.

\bibitem[LL07]{lussier2007cap}
Y.A. Lussier and Y.~Liu.
\newblock {Computational Approaches to Phenotyping: High-Throughput Phenomics}.
\newblock {\em Proceedings of the American Thoraic Society}, 4(1):18, 2007.

\bibitem[LLC{\etalchar{+}}03]{liu2003gtf}
M.~Liu, J.L. Leibowitz, D.A. Clark, M.~Mendicino, Q.~Ning, J.W. Ding,
  C.~D'Abreo, L.~Fung, P.A. Marsden, and G.A. Levy.
\newblock {Gene transcription of fgl2 in endothelial cells is controlled by
  Ets-1 and Oct-1 and requires the presence of both Sp1 and Sp3}.
\newblock {\em European Journal of Biochemistry}, 270(10):2274--2286, 2003.

\bibitem[LMA06]{liu2006ige}
Z.~Liu, K.~Maas, and T.M. Aune.
\newblock {Identification of gene expression signatures in autoimmune disease
  without the influence of familial resemblance}.
\newblock {\em Human Molecular Genetics}, 15(3):501--509, 2006.

\bibitem[LOT{\etalchar{+}}99]{lerebours1999fdm}
F.~Lerebours, S.~Olschwang, B.~Thuille, A.~Schmitz, P.~Fouchet, B.~Buecher,
  N.~Martinet, F.~Galateau, and G.~Thomas.
\newblock {Fine deletion mapping of chromosome 8p in non-small-cell lung
  carcinoma}.
\newblock {\em International Journal of Cancer}, 81(6), 1999.

\bibitem[MBBAC05]{mestiri2005fpt}
S.~Mestiri, N.~Bouaouina, S.~Ben~Ahmed, and L.~Chouchane.
\newblock {A functional polymorphism of the tumor necrosis factor receptor-II
  gene associated with the survival and relapse prediction of breast
  carcinoma}.
\newblock {\em Cytokine}, 30(4):182--187, 2005.

\bibitem[MERS96]{medvedev1996drt}
A.E. Medvedev, T.~Espevik, G.~Ranges, and A.~Sundan.
\newblock {Distinct roles of the two tumor necrosis factor (TNF) receptors in
  modulating TNF and lymphotoxin alpha effects}.
\newblock {\em Journal of Biological Chemistry}, 271(16):9778, 1996.

\bibitem[MFG{\etalchar{+}}03]{matys2003trt}
V.~Matys, E.~Fricke, R.~Geffers, E.~Gossling, M.~Haubrock, R.~Hehl,
  K.~Hornischer, D.~Karas, AE~Kel, OV~Kel-Margoulis, et~al.
\newblock {TRANSFAC (R): transcriptional regulation, from patterns to
  profiles}.
\newblock {\em Nucleic Acids Research}, 31(1):374, 2003.

\bibitem[MFK{\etalchar{+}}02]{motamed2002ips}
K.~Motamed, S.E. Funk, H.~Koyama, R.~Ross, E.W. Raines, and E.H. Sage.
\newblock {Inhibition of PDGF-stimulated and matrix-mediated proliferation of
  human vascular smooth muscle cells by SPARC is independent of changes in cell
  shape or cyclin-dependent kinase inhibitors}.
\newblock {\em Journal of cellular biochemistry}, 84(4), 2002.

\bibitem[MLH{\etalchar{+}}06]{miyamoto2006mpm}
A.~Miyamoto, R.~Lau, P.W. Hein, J.M. Shipley, and G.~Weinmaster.
\newblock {Microfibrillar proteins MAGP-1 and MAGP-2 induce Notch1
  extracellular domain dissociation and receptor activation}.
\newblock {\em Journal of Biological Chemistry}, 281(15):10089, 2006.

\bibitem[MR03]{xu2003gsf}
Xu~M. and Setiono R.
\newblock {Gene selection for cancer classification using a hybrid of
  univariate and multivariate feature selection methods}.
\newblock {\em Applied Genomics and Proteomics}, 2:79--91, 2003.

\bibitem[MVM{\etalchar{+}}04]{menendez2004nct}
J.A. Menendez, L.~Vellon, I.~Mehmi, P.K. Teng, D.W. Griggs, and R.~Lupu.
\newblock {A novel CYR61-triggered ``CYR61-$\alpha$v$\beta$3 integrin loop''
  regulates breast cancer cell survival and chemosensitivity through activation
  of ERK1/ERK2 MAPK signaling pathway}.
\newblock {\em Oncogene}, 24(5):761--779, 2004.

\bibitem[NASL01]{notterman2001tge}
D.A. Notterman, U.~Alon, A.J. Sierk, and A.J. Levine.
\newblock {Transcriptional gene expression profiles of colorectal adenoma,
  adenocarcinoma, and normal tissue examined by oligonucleotide arrays}, 2001.

\bibitem[NRT{\etalchar{+}}95]{neish1995eir}
AS~Neish, MA~Read, D.~Thanos, R.~Pine, T.~Maniatis, and T.~Collins.
\newblock {Endothelial interferon regulatory factor 1 cooperates with NF-kappa
  B as a transcriptional activator of vascular cell adhesion molecule 1}.
\newblock {\em Molecular and Cellular Biology}, 15(5):2558--2569, 1995.

\bibitem[OHB{\etalchar{+}}02]{ortego2002hmc}
M.~Ortego, A.G. Hern{\'a}ndez, C.~Bustos, L.M. Blanco-Colio, M.A.
  Hern{\'a}ndez-Presa, J.~Tu{\~n}{\'o}n, and J.~Egido.
\newblock {3-Hydroxy-3-methylglutaryl coenzyme A reductase inhibitors increase
  the binding activity and nuclear level of Oct-1 in mononuclear cells}.
\newblock {\em European journal of pharmacology}, 448(2-3):113--121, 2002.

\bibitem[OHB08]{oti2008pc}
M.~Oti, M.A. Huynen, and H.G. Brunner.
\newblock {Phenome connections}.
\newblock {\em Trends in Genetics}, 24(3):103--106, 2008.

\bibitem[OMF04]{ottaviani2004ihf}
E.~Ottaviani, D.~Malagoli, and A.~Franchini.
\newblock {Invertebrate humoral factors: cytokines as mediators of cell
  survival}.
\newblock {\em Invertebrate Cytokines and the Phylogeny of Immunity: Facts and
  Paradoxes}, page~1, 2004.

\bibitem[PAF{\etalchar{+}}03]{pupa2003iap}
S.M. Pupa, S.W. Argraves, S.~Forti, P.~Casalini, V.~Berno, R.~Agresti,
  P.~Aiello, A.~Invernizzi, P.~Baldassari, W.~Otwal, et~al.
\newblock {Immunological and pathobiological roles of fibulin-1 in breast
  cancer}.
\newblock {\em Oncogene}, 23(12):2153--2160, 2003.

\bibitem[PBG{\etalchar{+}}00]{pettersson2000ccl}
M.~Pettersson, M.~Bessonova, H.F. Gu, L.C. Groop, and J.I. J{\"o}nsson.
\newblock {Characterization, chromosomal localization, and expression during
  hematopoietic differentiation of the gene encoding Arl6ip,
  ADP-ribosylation-like factor-6 interacting protein (ARL6)}.
\newblock {\em Genomics}, 68(3):351--354, 2000.

\bibitem[PFD{\etalchar{+}}06]{polfliet2006rms}
M.M.J. Polfliet, B.O. Fabriek, W.P. Dani{\"e}ls, C.D. Dijkstra, and T.K.
  van~den Berg.
\newblock {The rat macrophage scavenger receptor CD163: expression, regulation
  and role in inflammatory mediator production}.
\newblock {\em Immunobiology}, 211(6-8):419--425, 2006.

\bibitem[RASB{\etalchar{+}}03]{rennstam2003pci}
K.~Rennstam, M.~Ahlstedt-Soini, B.~Baldetorp, P.O. Bendahl, A.~Borg, R.~Karhu,
  M.~Tanner, M.~Tirkkonen, and J.~Isola.
\newblock {Patterns of chromosomal imbalances defines subgroups of breast
  cancer with distinct clinical features and prognosis. A study of 305 tumors
  by comparative genomic hybridization}, 2003.

\bibitem[RBR{\etalchar{+}}02]{rhodes2002mam}
D.R. Rhodes, T.R. Barrette, M.A. Rubin, D.~Ghosh, and A.M. Chinnaiyan.
\newblock {Meta-Analysis of Microarrays Interstudy Validation of Gene
  Expression Profiles Reveals Pathway Dysregulation in Prostate Cancer 1},
  2002.

\bibitem[RLIA{\etalchar{+}}92]{raines1992egs}
EW~Raines, TF~Lane, ML~Iruela-Arispe, R.~Ross, and EH~Sage.
\newblock {The extracellular glycoprotein SPARC interacts with platelet-derived
  growth factor (PDGF)-AB and-BB and inhibits the binding of PDGF to its
  receptors}.
\newblock {\em Proceedings of the National Academy of Sciences},
  89(4):1281--1285, 1992.

\bibitem[RLS{\etalchar{+}}04]{ross2004bcb}
J.S. Ross, G.P. Linette, J.~Stec, E.~Clark, M.~Ayers, N.~Leschly, W.F. Symmans,
  G.N. Hortobagyi, and L.~Pusztai.
\newblock {Breast cancer biomarkers and molecular medicine: part II}.
\newblock {\em erm}, 4(2):169--188, 2004.

\bibitem[RYS{\etalchar{+}}04]{rhodes2004lsm}
D.R. Rhodes, J.~Yu, K.~Shanker, N.~Deshpande, R.~Varambally, D.~Ghosh,
  T.~Barrette, A.~Pandey, and A.M. Chinnaiyan.
\newblock {Large-scale meta-analysis of cancer microarray data identifies
  common transcriptional profiles of neoplastic transformation and
  progression}.
\newblock {\em Proceedings of the National Academy of Sciences},
  101(25):9309--9314, 2004.

\bibitem[SB88]{salton1988twa}
G.~Salton and C.~Buckley.
\newblock {Term-weighting approaches in automatic text retrieval}.
\newblock {\em Information Processing and Management: an International
  Journal}, 24(5):513--523, 1988.

\bibitem[SBB{\etalchar{+}}05]{seol2005epa}
MB~Seol, JJ~Bong, M.~Baik, et~al.
\newblock {Expression profiles of apoptosis genes in mammary epithelial cells}.
\newblock {\em Molecules and Cells}, 2005.

\bibitem[SC03]{sevenet2003dmc}
N.~Sevenet and O.~Cussenot.
\newblock {DNA microarrays in clinical practice: past, present, and future.}
\newblock {\em Clinical and experimental medicine}, 3(1):1, 2003.

\bibitem[SFKR04]{segal2004mms}
E.~Segal, N.~Friedman, D.~Koller, and A.~Regev.
\newblock {A module map showing conditional activity of expression modules in
  cancer}.
\newblock {\em Nature genetics}, 36:1090--1098, 2004.

\bibitem[SFR{\etalchar{+}}02]{singh2002gec}
D.~Singh, P.G. Febbo, K.~Ross, D.G. Jackson, J.~Manola, C.~Ladd, P.~Tamayo,
  A.A. Renshaw, A.V. D'Amico, J.P. Richie, et~al.
\newblock {Gene expression correlates of clinical prostate cancer behavior}.
\newblock {\em Cancer cell}, 1(2):203--209, 2002.

\bibitem[SGC{\etalchar{+}}06]{silva2006sva}
HC~Silva, F.~Garcao, EC~Coutinho, CF~De~Oliveira, and FJ~Regateiro.
\newblock {Soluble VCAM-1 and E-selectin in breast cancer: relationship with
  staging and with the detection of circulating cancer cells}.
\newblock {\em Neoplasma}, 53(6):538--543, 2006.

\bibitem[SHG04]{smit2004ro}
AFA Smit, R.~Hubley, and P.~Green.
\newblock {RepeatMasker Open-3.0}.
\newblock {\em Institute for Systems Biology. http://www. repeatmasker. org
  (January 10, 2007)}, 2004.

\bibitem[SKFW03]{steuer2003oai}
R.~Steuer, J.~Kurths, O.~Fiehn, and W.~Weckwerth.
\newblock {Observing and interpreting correlations in metabolomic networks},
  2003.

\bibitem[SKW{\etalchar{+}}06]{seitz2006gbd}
S.~Seitz, E.~Korsching, J.~Weimer, A.~Jacobsen, N.~Arnold, A.~Meindl,
  W.~Arnold, D.~Gustavus, C.~Klebig, I.~Petersen, et~al.
\newblock {Genetic background of different cancer cell lines influences the
  gene set involved in chromosome 8 mediated breast tumor suppression}.
\newblock {\em Genes, Chromosomes and Cancer}, 45(6), 2006.

\bibitem[SOG{\etalchar{+}}03]{syder2003ipc}
A.J. Syder, J.D. Oh, J.L. Guruge, D.~O'Donnell, M.~Karlsson, J.C. Mills, B.M.
  Bjorkholm, and J.I. Gordon.
\newblock {The impact of parietal cells on Helicobacter pylori tropism and host
  pathology: an analysis using gnotobiotic normal and transgenic mice}.
\newblock {\em Proceedings of the National Academy of Sciences},
  100(6):3467--3472, 2003.

\bibitem[SR04]{sriraman2004cge}
V.~Sriraman and J.A.S. Richards.
\newblock {Cathepsin L gene expression and promoter activation in rodent
  granulosa cells}.
\newblock {\em Endocrinology}, 145(2):582--591, 2004.

\bibitem[SRJ{\etalchar{+}}98]{schwachtgen1998oit}
J.L. Schwachtgen, J.E. Remacle, N.~Janel, R.~Brys, D.~Huylebroeck, D.~Meyer,
  and D.~Kerbiriou-Nabias.
\newblock {Oct-1 is involved in the transcriptional repression of the von
  Willebrand factor gene promoter}.
\newblock {\em Blood}, 92(4):1247, 1998.

\bibitem[SRT{\etalchar{+}}02]{shipp2002dlb}
M.A. Shipp, K.N. Ross, P.~Tamayo, A.P. Weng, J.L. Kutok, R.C.T. Aguiar,
  M.~Gaasenbeek, M.~Angelo, M.~Reich, G.S. Pinkus, et~al.
\newblock {Diffuse large B-cell lymphoma outcome prediction by gene-expression
  profiling and supervised machine learning}.
\newblock {\em Nature medicine}, 8(1):68--74, 2002.

\bibitem[SWF{\etalchar{+}}00]{seitz2000rdm}
S.~Seitz, S.~Werner, J.~Fischer, A.~Nothnagel, PM~Schlag, and S.~Scherneck.
\newblock {Refined deletion mapping in sporadic breast cancer at chromosomal
  region 8p12-p21 and association with clinicopathological parameters}.
\newblock {\em European Journal of Cancer}, 36(12):1507--1513, 2000.

\bibitem[THC{\etalchar{+}}99]{tavazoie1999sdg}
S.~Tavazoie, J.D. Hughes, M.J. Campbell, R.J. Cho, and G.M. Church.
\newblock {Systematic determination of genetic network architecture}.
\newblock {\em Nature genetics}, 22:281--285, 1999.

\bibitem[TMAH05]{takayama2005ldl}
Y.~Takayama, P.~May, R.G.W. Anderson, and J.~Herz.
\newblock {Low density lipoprotein receptor-related protein 1 (LRP1) controls
  endocytosis and c-CBL-mediated ubiquitination of the platelet-derived growth
  factor receptor $\{$beta$\}$(PDGFR $\{$beta$\}$)}.
\newblock {\em Journal of Biological Chemistry}, 280(18):18504, 2005.

\bibitem[TSE{\etalchar{+}}07]{tamayo2007mpc}
P.~Tamayo, D.~Scanfeld, B.L. Ebert, M.A. Gillette, C.W.M. Roberts, and J.P.
  Mesirov.
\newblock {Metagene projection for cross-platform, cross-species
  characterization of global transcriptional states}.
\newblock {\em Proceedings of the National Academy of Sciences}, 104(14):5959,
  2007.

\bibitem[UK05]{ustach2005pdg}
C.V. Ustach and H.R.C. Kim.
\newblock {Platelet-derived growth factor D is activated by urokinase
  plasminogen activator in prostate carcinoma cells}.
\newblock {\em Molecular and Cellular Biology}, 25(14):6279--6288, 2005.

\bibitem[Vap00]{vapnik2000nsl}
V.N. Vapnik.
\newblock {\em {The nature of statistical learning theory}}.
\newblock springer, 2000.

\bibitem[WBD{\etalchar{+}}01]{west2001pcs}
M.~West, C.~Blanchette, H.~Dressman, E.~Huang, S.~Ishida, R.~Spang, H.~Zuzan,
  J.A. Olson~Jr, J.R. Marks, and J.R. Nevins.
\newblock {Predicting the clinical status of human breast cancer by using gene
  expression profiles}.
\newblock {\em Proceedings of the National Academy of Sciences}, 98(20):11462,
  2001.

\bibitem[WDJB{\etalchar{+}}05]{watkins2005ils}
G.~Watkins, A.~Douglas-Jones, R.~Bryce, R.~E~Mansel, and W.G. Jiang.
\newblock {Increased levels of SPARC (osteonectin) in human breast cancer
  tissues and its association with clinical outcomes}.
\newblock {\em Prostaglandins, Leukotrienes \& Essential Fatty Acids},
  72(4):267--272, 2005.

\bibitem[WHD{\etalchar{+}}02]{wu2002lsp}
L.F. Wu, T.R. Hughes, A.P. Davierwala, M.D. Robinson, R.~Stoughton, and S.J.
  Altschuler.
\newblock {Large-scale prediction of Saccharomyces cerevisiae gene function
  using overlapping transcriptional clusters}.
\newblock {\em nature genetics}, 31:255--265, 2002.

\bibitem[Wil05]{wilcox2005ire}
R.R. Wilcox.
\newblock {\em {Introduction to robust estimation and hypothesis testing}}.
\newblock Academic Press, 2005.

\bibitem[XFZ01]{xiong2001bif}
M.~Xiong, X.~Fang, and J.~Zhao.
\newblock {Biomarker identification by feature wrappers}, 2001.

\bibitem[XKNI{\etalchar{+}}08]{xu2008iac}
M.~Xu, M.C. Kao, J.~Nunez-Iglesias, J.~Nevins, M.~West, and X.J. Zhou.
\newblock {An integrative approach to characterize disease-specific pathways
  and their coordination: a case study in cancer}.
\newblock {\em BMC genomics}, 9(Suppl 1):S12, 2008.

\bibitem[XZZ08]{xu2008sim}
M.~Xu, M.~Zhu, and L.~Zhang.
\newblock {A stable iterative method for refining discriminative gene
  clusters}.
\newblock {\em BMC genomics}, 9(Suppl 2):S18, 2008.

\bibitem[YKL{\etalchar{+}}96]{yaremko1996lhs}
M.L. Yaremko, C.~Kutza, J.~Lyzak, R.~Mick, W.M. Recant, and C.A. Westbrook.
\newblock {Loss of heterozygosity from the short arm of chromosome 8 is
  associated with invasive behavior in breast cancer}.
\newblock {\em Genes, Chromosomes and Cancer}, 16(3), 1996.

\bibitem[YMH{\etalchar{+}}07]{yan2007gba}
X.~Yan, M.R. Mehan, Y.~Huang, M.S. Waterman, P.S. Yu, and X.J. Zhou.
\newblock {A graph-based approach to systematically reconstruct human
  transcriptional regulatory modules}.
\newblock {\em Bioinformatics}, 23(13):i577, 2007.

\bibitem[ZBC{\etalchar{+}}06]{zymek2006rpd}
P.~Zymek, M.~Bujak, K.~Chatila, A.~Cieslak, G.~Thakker, M.L. Entman, and N.G.
  Frangogiannis.
\newblock {The role of platelet-derived growth factor signaling in healing
  myocardial infarcts}.
\newblock {\em Journal of the American College of Cardiology},
  48(11):2315--2323, 2006.

\bibitem[ZKH{\etalchar{+}}05]{zhou2005faa}
X.J. Zhou, M.C.J. Kao, H.~Huang, A.~Wong, J.~Nunez-Iglesias, M.~Primig, O.M.
  Aparicio, C.E. Finch, T.E. Morgan, and W.H. Wong.
\newblock {Functional annotation and network reconstruction through
  cross-platform integration of microarray data}.
\newblock {\em Nature Biotechnology}, 23(2):238--243, 2005.

\bibitem[ZLS{\etalchar{+}}06]{zhang2006rsf}
X.~Zhang, X.~Lu, Q.~Shi, X.~Xu, H.E. Leung, L.N. Harris, J.D. Iglehart,
  A.~Miron, J.S. Liu, and W.H. Wong.
\newblock {Recursive SVM feature selection and sample classification for
  mass-spectrometry and microarray data}.
\newblock {\em BMC bioinformatics}, 7(1):197, 2006.

\bibitem[ZW08]{zhu2008pca}
M.~Zhu and Q.~Wu.
\newblock {A Parallel Computing Approach to Decipher Transcription Network for
  Large-scale Microarray Datasets}.
\newblock {\em BMC Genomics}, 9(Suppl 1):S5, 2008.

\end{thebibliography}
\end{document}